\documentclass[symmetry,review,accept,pdftex,moreauthors]{Definitions/mdpi} 

\firstpage{1} 
\pubvolume{15}
\issuenum{5}
\articlenumber{978}
\pubyear{2023}
\copyrightyear{2023}
 \externaleditor{Academic Editor: {Giovanni Modanese} 
}
\datereceived{26 March 2023} 
\daterevised{20 April 2023} 
\dateaccepted{21 April 2023} 
\datepublished{25 April 2023} 
\hreflink{\textls[-35]{https://doi.org/10.3390/sym15050978}} 

\usepackage{amsthm}
\usepackage{mathtools}
\usepackage{xcolor}
\usepackage{graphicx}
\usepackage{lipsum}
\usepackage{tensor}

\Title{Lorentz Violation in Finsler Geometry}

\TitleCitation{Lorentz Violation in Finsler Geometry}

\Author{ {Jie} 
 Zhu $^{1}$\orcidA{} and Bo-Qiang Ma $^{1,2,3,*}$\orcidB{} }

\AuthorNames{Jie Zhu and Bo-Qiang Ma}

\AuthorCitation{Zhu, J.; Ma, B.-Q.}

\address{%
$^{1}$ \quad School of Physics, Peking University, Beijing 100871, China; jiezhu@pku.edu.cn~(J.Z.); mabq@pku.edu.cn~(B.-Q.M.)\\
$^{2}$ \quad Center for High Energy Physics, Peking University, Beijing 100871, China\\
$^{3}$ \quad Collaborative Innovation Center of Quantum Matter,  {Beijing,} 
 China}

\corres{Correspondence: mabq@pku.edu.cn}

\abstract{Lorentz invariance is one of the foundations of modern physics; however, Lorentz violation may happen from the perspective of quantum gravity, and plenty of studies on Lorentz violation have arisen in recent years. As a good tool to explore Lorentz violation, Finsler geometry is a natural and fundamental generalization of Riemann geometry. The Finsler structure depends on both coordinates and velocities. 
Here, we simply introduce the mathematics of Finsler geometry.
We review the connection between modified dispersion relations and Finsler geometries and discuss the physical influence from Finsler geometry.
We review the connection between Finsler geometries and theories of Lorentz violation, such as the doubly special relativity, 
the standard-model extension, 
and the very special relativity. 
}

\begin{document}
\section{Introduction} \label{sec:introduction}

As a basic symmetry of spacetime, Lorentz invariance is a basic assumption of Einstein's relativity,
and now it is the foundation of general relativity and quantum field theory in modern physics.
However, in quantum gravity (QG)~\cite{qg}, Lorentz violation (LV) may happen at the Planck scale ($E_{\rm Pl}\simeq 1.22\times 10^{19}~\rm{GeV}$).
There are many kinds of theoretical models including LV effects, 
including the QG theories such as string theory~\cite{string,Chengyi2021,Chengyi2022,Chengyi2023} and loop quantum gravity~\cite{Rovelli2008, Ashtekar2021, lihao2023},
spacetime structure theories such as doubly special relativity (DSR)~\cite{DSR1-1, DSR1-2, DSR2-1, DSR2-2} and very special relativity (VSR)~\cite{vsr}, 
and the effective theory with extra terms, such as the standard-model extension (SME)~\cite{cpt-sm, lv-sm, glv-sm}.
For a recent review on Lorentz invariance violation, see~\cite{HeMa}.

Among many LV theories, modified dispersion relations (MDRs) are introduced, 
and for energy $E\ll E_{\rm Pl}$, the MDR can be expressed at the leading order as
\begin{equation}
    E^2=m^2+p^2+\alpha p^{n+2},
\end{equation}
where $n$ is the broken order here, and $\alpha$ is a parameter with mass dimension and $[\alpha]=-n$.
In 2007,  {Girelli} 
  {et~al.}~found the relations between MDRs and Finsler geometries, 
and in the same year, Gibbons  {et~al.}~found that the general VSR is Finsler geometry.
Since then, Finsler geometry has drawn more attention to LV studies, for example, the application of Randers--Finsler geometry on LV~\cite{ChangZ2008, ChangZ2009}.

Finsler geometry is the generalization of Riemann geometry and is named after Paul Finsler who studied it in his doctoral thesis in 1917.
Finsler geometries have many applications in different physics domains.
Since Riemann geometry describes gravity, it is natural to study gravitation via Finsler geometry to pursue new physics.
The gravitation theory in Berwald–Finsler spacetime is built up in Ref.~\cite{LiX2010},
and the vacuum field equation in Finsler spacetime is present in Ref.~\cite{Pfeifer2012}.
The modification of Newtonian dynamics is also considered in Finsler geometry~\cite{LiX2013}.
In addition, Finsler geometries are applied to cosmology problems, such as dark matter~\cite{ChangZ2008-2, Saridakis2013}, dark energy~\cite{ChangZ2009-2}, bounce cosmology~\cite{Saridakis2019-1},
and anisotropy of the universe~\cite{ChangZ2014, ChangZ2014-2}, 
and Finsler-like theories~\cite{Saridakis2019-2, Saridakis2021} are also applied to cosmology.
Gravitational waves are also studied in Finsler spacetime~\cite{LiX2013-2}.
Even neutrino oscillations have a realization in Finsler spacetime~\cite{Antonelli2018}.
One can conclude from the above examples that Finsler geometry has an infinite number of degrees of freedom to choose from, 
which can be modified at will to meet different requirements in different physics domains.
It is not reasonable to introduce all of the applications of different kinds of Finsler geometries in a short review.
In this review, we focus on the applications of Finsler geometry on Lorentz invariance violation.

This review is organized as follows: 
Section~\ref{sec:finsler} briefly introduces the mathematics of Finsler geometry.
Section~\ref{sec:mdrF} introduces the connection between MDRs and Finsler geometries.
Section~\ref{sec:lvF} introduces the connection between LV theories and Finsler geometries, including DSR, SME, and VSR.
Section~\ref{sec:discussion} reviews and discusses the topic of this review.

\section{Introduction to Finsler Geometry}\label{sec:finsler}

\subsection{A First Glance at Finsler Geometry}
Essentially, a Finsler manifold is a manifold $M$ where each tangent space is equipped with a Minkowski norm, that is, a norm that is not necessarily induced by an inner product. 
The metric of a Finsler manifold depends on not only the points of $M$, but also the directions in the tangent bundle of $M$.

For a manifold $M$, denote by $T_x M$ the tangent space at $x \in M$, and by $T M$ the tangent bundle of $M$. Each element of $T M$ has the form $(x, y)$, where $x \in M$ and $y \in T_x M$. The natural projection $\pi: T M \rightarrow M$ is given by $\pi(x, y) \equiv x$.
A Finsler structure or Finsler norm of $M$ is a function
\begin{equation}
    F: T M \rightarrow[0, \infty)
\end{equation}
with the following properties:
\begin{enumerate}
\item Regularity: $\mathrm{F}$ is $C^{\infty}$ on the entire slit tangent bundle $TM_0 :=T M \backslash \{0\}$.
\item Positive homogeneity : $F(x, \lambda y)=\lambda F(x, y)$ for all $\lambda>0$.
\item Strong convexity: The $n \times n$ Hessian matrix\vspace{-6pt}
\begin{equation}\label{eq:gij}
    g_{i j} :=\left(\frac{1}{2} F^2\right)_{y^i y^j}
\end{equation}
is positive-definite at every point of $T M_0$, where we use the notation ()$_{y^i}=$ $\frac{\partial}{\partial y^i}()$.
\end{enumerate}

$g_{ij}$  {in} 
 Equation~(\ref{eq:gij}) is called the Finsler metric.
From Euler's homogeneous function theorem, the Finsler norm $F$ has attributions such that
\begin{equation}\label{eq:homo}
    y^iF_{y^i}=F,\quad y^i F_{y^i y^j}=0,\quad F^2=g_{ij}(x,y)y^iy^j.
\end{equation}

Finsler geometry has its genesis in integrals of the form
\begin{equation}\label{eq:int}
    L_F(C)=\int_C F\left(x^1, \cdots, x^n ; \frac{d x^1}{d \tau}, \cdots, \frac{d x^n}{d \tau}\right) d \tau,
\end{equation}
and its geometric meaning is the distance between two points in the Finsler manifold through a certain path $C$.
Given a manifold $M$ and a Finsler structure $F$ on $T M$, the pair $(M, F)$ is called a Finsler manifold. The Finsler structure $F$ is a function of $\left(x^i, y^i\right)$. In the case of $g_{ij}$, depending on $x^i$ only, the Finsler manifold reduces to Riemannian manifold.

To make things easier understood by physicists, here are some examples.
As mentioned above, Riemann geometry is a special Finsler geometry. 
Riemann geometry can be described by the Riemann metric
\begin{equation}
    ds^2=g_{\mu\nu}(x) dx^\mu dx^\nu, 
\end{equation}
and, here, $g_{\mu\nu}(x)$ depends only on the point $x$.
In the language of Finsler geometry, the Finsler structure of the above Riemann metric is
\begin{equation}
    F=\sqrt{g_{\mu\nu}(x) y^\mu y^\nu},
\end{equation}
where $y^\mu=dx^\mu/d\tau$.
A kind of special Finsler geometry is of Randers type, and its Finsler norm is modified from Riemann metric as
\begin{equation}
    F=\sqrt{a_{\mu\nu}(x) y^\mu y^\nu}+b_\mu(x) y^\mu=\alpha(x,y)+\beta(x,y),
\end{equation}
where $\alpha(x,y)=\sqrt{a_{\mu\nu}(x) y^\mu y^\nu}$ and $\beta(x,y)=b_\mu(x) y^\mu$.
Generally, any form of 
\begin{equation}
    F=\alpha(x,y)\phi(s), \quad s=\frac{\beta(x,y)}{\alpha(x,y)},
\end{equation}
is a Finsler structure.
Another example is that in a universe where the Pythagorean theorem is modified from $c^2=a^2+b^2$ to $c^4=a^4+b^4$, the length element in a two-dimensional plane is $ds^4=dx_1^4+dx_2^4$; thus, the Finsler norm in this plane is
\begin{equation}
    F=(y_1^4+y_2^4)^\frac{1}{4}.
\end{equation}
 {We can} 
 calculate the Finsler metric from Equation~(\ref{eq:gij}) as
\begin{equation}
    g_{ij}(x,y)=\begin{pmatrix}
\frac{y_1^6+3 y_1^2 y_2^4}{\left(y_1^4+y_2^4\right)^{3/2}} & -\frac{2 y_1^3 y_2^3}{\left(y_1^4+y_2^4\right)^{3/2}} \\
-\frac{2 y_1^3 y_2^3}{\left(y_1^4+y_2^4\right)^{3/2}} & \frac{3 y_1^4 y_2^2+y_2^6}{\left(y_1^4+y_2^4\right)^{3/2}} 
\end{pmatrix}.
\end{equation}
 {In this }example, the Finsler metric depends only on $y$.
We can identify the differences between Finsler metrics and Riemann metrics in two ways: whether the square of the length element is a quadratic form of $y$, or whether the metric $g_{ij}(x,y)$ contains $y$.

To describe the ``1  {+} 
 3'' spacetime, instead of Finsler geometry we turn to pseudo-Finsler geometry, 
similar to the relation between Riemann geometry and pseudo-Riemann geometry that describes general relativity.
To simplify the description, in the following context, we do not distinguish between Finsler geometry and pseudo-Finsler geometry; we call them Finsler geometry.

\subsection{Mathematical Concepts of Finsler Geometry}

Finsler geometry contains analogs for many of the natural objects in Riemannian geometry, but their formulas are very different. For example, the Christoffel symbol in Riemann geometry plays the role in the connection and can be expressed as
\begin{equation}\label{eq:chris}
    \gamma_{j k}^i = \frac{g^{i l}}{2}\left(\frac{\partial g_{l j}}{\partial x^k}+\frac{\partial g_{l k}}{\partial x^j}-\frac{\partial g_{j k}}{\partial x^l}\right), 
\end{equation}
where $g^{i l}$ is the inverse matrix of $g_{i l}$. In Finsler geometry, $\gamma_{j k}^i$ in Equation~(\ref{eq:chris}) is less important and is called the \emph{ {formal Christoffel symbol.} 
}
Since this review is for physicists to quickly understand Finsler geometry, here we just review some major concepts in Finsler geometry quickly. For detailed discussions see Ref.~\cite{textbook}. 

The \emph{ {Hilbert form}} is a 1-form defined as 
\begin{equation}
    \omega := F_{y^i} dx^i = \frac{y^j}{F}g_{ji}dx^i,
\end{equation}
and its dual vector field is
\begin{equation}
    l=l^i\frac{\partial}{\partial x^i}:=\frac{y^i}{F}\frac{\partial}{\partial x^i}.
\end{equation}
 {The} Hilbert form is invariant under coordinate transformations. Assume that $(\Tilde{x},\Tilde{y})$ is another local coordination system, thus
\begin{equation}
    \Tilde{y}^i=\frac{\partial \Tilde{x}^i}{\partial x^j}y^j, F_{y^i}=\frac{\partial \Tilde{x}^j}{\partial x^i}F_{\Tilde{y}^j}, 
\end{equation}
so 
\begin{equation}
    \omega = F_{y^i} dx^i = F_{\Tilde{y}^i} d\Tilde{x}^i.
\end{equation}
 {From} Equation~(\ref{eq:homo}), the distance integration in Equation~(\ref{eq:int}) can also be expressed as
\begin{equation}
    L_F(C)=\int_C \omega.
\end{equation}

The \emph{Cartan tensor} is defined as
\begin{equation}
\begin{split}
    &A_{ijk}:=\frac{F}{2}\frac{\partial g_{ij}}{\partial y^k}=\frac{F}{4}(F^2)_{y^iy^jy^k},\\
    &A:=A_{ijk}dx^i\otimes dx^j \otimes dx^k.
\end{split}
\end{equation}
Obviously, a Finsler manifold is a Riemann manifold if and only if $A\equiv 0$.
The Cartan tensor is a symmetric covariant tensor with an attribute
\begin{equation}
    y^iA_{ijk}=y^jA_{ijk}=y^kA_{ijk}=0.
\end{equation}

As a vector field on $TM$, the transformation formula of $\frac{\partial}{\partial x^i}$ is
\begin{equation}
    \frac{\partial}{\partial \Tilde{x}^i}=\frac{\partial x^k}{\partial \Tilde{x}^i} \frac{\partial}{\partial x^k}+\frac{\partial^2x^k}{\partial \Tilde{x}^i\partial \Tilde{x}^j}\Tilde{y}^j\frac{\partial}{\partial y^k}.
\end{equation}
To obtain the same transformation law as a tensor, we need to replace $\frac{\partial}{\partial x^i}$ by $\frac{\delta}{\delta x^i}$ as
\begin{equation}
    \frac{\delta}{\delta x^i}:=\frac{\partial}{\partial x^i}-N^j_i\frac{\partial}{\partial y^j},
\end{equation}
where
\begin{equation}
    N^j_i:=\gamma^j_{ik}y^k-g^{jl}\frac{A_{ikl}}{F}\gamma^k_{pq}y^py^q
\end{equation}
is connection, and $\gamma^i_{jk}$ is the formal Christoffel symbol in Equation~(\ref{eq:chris}).
One can easily check 
\begin{equation}
    \frac{\delta}{\delta \Tilde{x}^i}=\frac{\partial x^k}{\partial \Tilde{x}^i}\frac{\delta}{\delta x^k}.
\end{equation}
Denote by $T(TM_0)$ the tangent bundle of $TM_0$, and $T(TM_0)$ can be divided into the horizontal part $HTM$ spanned by $\{\frac{\delta}{\delta x^i}\}$ and the vertical part $VTM$ spanned by $\{\frac{\partial}{\partial y^i}\}$. Their dual bases are $\{dx^i\}$ and $\{\delta y^i\}$, where
\begin{equation}
    \delta y^i:= dy^i+N^i_jdx^j.
\end{equation}
In other words,
\begin{equation}
    \begin{split}
        &T(TM_0) =HTM\oplus VTM=\text{span}\{\frac{\delta}{\delta x^i}\}\oplus\text{span}\{\frac{\partial}{\partial y^i}\},\\
        &T^*(TM_0)=H^*TM\oplus V^*TM=\text{span}\{dx^i\}\oplus\text{span}\{\delta y^i\}.
    \end{split}
\end{equation}
Since Finsler geometry is discussed on $TM$, the division above is important and shows the differences with Riemann geometry.


\emph{Chern connections} are connection 1-forms $\omega^i_j$ with attributes torsion freeness
\begin{equation}
    d x^j \wedge \omega_j^i=0
\end{equation}
and almost $g$-compatibility
\begin{equation}
    d g_{i j}-g_{k j} \omega_i^k-g_{i k} \omega_j^k=2 \frac{A_{i j s}}{F} \delta y^s.
\end{equation}
The Chern 1-form can be expressed as
\begin{equation}
    \omega^i_j=\Gamma ^i_{jk}dx^k,
\end{equation}
where $\Gamma ^i_{jk}$ is the \emph{Christoffel symbol (or the connection coefficients) of the Chern connection} with an attribute $\Gamma ^i_{jk}=\Gamma ^i_{kj}$.
The Christoffel symbol of the Chern connection can be expressed as
\begin{equation}
\begin{split}
    \Gamma^l_{ij}&=\frac{1}{2}g^{lk}\left(\frac{\delta g_{ik}}{\delta x^j}+\frac{\delta g_{jk}}{\delta x^i}-\frac{\delta g_{ij}}{\delta x^k}\right)\\
    &=\gamma^l_{ij}-\frac{g^{lk}}{F}(A_{jkp}N^p_i+A_{kip}N^p_j-A_{ijp}N^p_k).
\end{split}
\end{equation}

For any tensor on $TM_0$, we can define its covariant derivative via Chern connections. 
Since $T^*(TM_0)$ can be divided into two parts, the covariant derivatives can be also divided into horizontal parts and vertical parts.
For example, for a $(1,1)$-tensor $T=\tensor{T}{^i_j}\frac{\partial}{\partial x^i}\otimes dx^j$,
its covariant derivative can be defined as
\begin{equation}
    \nabla \tensor{T}{^i_j}:= d\tensor{T}{^i_j}+\tensor{T}{^k_j}\omega^i_k-\tensor{T}{^i_k}\omega^k_j=\tensor{T}{^i_{j|k}}dx^k+\tensor{T}{^i_{j;k}}\delta y^k,
\end{equation}
where $\tensor{T}{^i_{j|k}}$ and $\tensor{T}{^i_{j;k}}$ are called \emph{horizontal covariant derivative} and \emph{vertical covariant derivative}, respectively. Their expressions are shown as follows:
\begin{equation}
\begin{split}
    \tensor{T}{^i_{j|k}}&=\frac{\partial \tensor{T}{^i_j}}{\partial x^k}+\tensor{T}{^s_j}\Gamma^i_{sk}-\tensor{T}{^i_s}\Gamma^s_{jk}-\tensor{T}{^i_{j;s}}N^s_k\\
    &=\frac{\delta \tensor{T}{^i_j}}{\delta x^k}+\tensor{T}{^s_j}\Gamma^i_{sk}-\tensor{T}{^i_s}\Gamma^s_{jk},\\
    \tensor{T}{^i_{j;k}}&=\frac{\partial \tensor{T}{^i_j}}{\partial y^k}.
\end{split}
\end{equation}

The \emph{geodesic spray coefficient} is $G^i$, given by
\begin{equation}
    G^i=\frac{1}{4}g^{ij}\left((F^2)_{y^j x^k} y^k -(F^2)_{x^j}\right).
\end{equation}
The geodesic equation for the Finsler manifold is given as
\begin{equation}\label{eq:geodesic}
    \frac{d^{2} x^i}{d \tau^{2}}+2 G^i=0. 
\end{equation}
The connections between geodesic spray coefficients and Christoffel symbols are
\begin{equation}
    G^i=\frac{1}{2}\gamma^i_{jk}y^jy^k=\frac{1}{2}\Gamma^i_{jk}y^jy^k, 
\end{equation}
so the geodesic equations in Finsler manifolds are the same as in Riemann manifolds.
The connections between $\Gamma^l_{ij}$, $N^i_j$ and $G^i$ are
\begin{equation}
    N^i_j=\Gamma^i_{jk}y^k=\frac{\partial G^i}{\partial y^j},
\end{equation}
thus $N^i_j$ only depends on $G^i$.

The curvature 2-forms of the Chern connection are
\begin{equation}
\begin{split}
    \Omega^i_j:&=d\omega^i_j-\omega^k_j\wedge \omega^i_k\\
    &=\frac{1}{2}\tensor{R}{_j^i_k_l}dx^k\wedge dx^l+\tensor{P}{_j^i_k_l}dx^k\wedge\delta y^l,
\end{split}
\end{equation}
where $R$ and $P$ are the  $hh-$ and $hv-$\emph{curvature tensors of the Chern connection}, respectively.
The expression of the $hh-$curvature tensor is
\begin{equation}
    \tensor{R}{_j^i_k_l}=\frac{\delta\Gamma^i_{jl}}{\delta x^k}-\frac{\delta\Gamma^i_{jk}}{\delta x^l}+\Gamma^i_{km}\Gamma^m_{jl}-\Gamma^i_{lm}\Gamma^m_{jk},
\end{equation}
and the $hh-$curvature tensor has attributes
\begin{equation}
\begin{split}
    &\tensor{R}{_j^i_k_l}+\tensor{R}{_j^i_l_k}=0,\\
    &\tensor{R}{_j^i_k_l}+\tensor{R}{_k^i_l_j}+\tensor{R}{_l^i_j_k}=0.
\end{split}
\end{equation}
The expression of the $hv-$curvature tensor is
\begin{equation}
    \tensor{P}{_j^i_k_l}=-\frac{\partial \Gamma^i_{jk}}{\partial y^l}.
\end{equation}
Here, we can see another difference between Riemann manifolds and Finsler manifolds: the curvature tensors have extra $hv$ parts in Finsler manifolds. 
The $hv-$curvature tensors vanish if and only if the Christoffel symbol of the Chern connection $\Gamma^i_{jk}=\Gamma^i_{jk}(x)$ is only the function of the point $x$ of the manifold $M$. 
In this case, we call $M$ a \emph{Berwald–Finsler manifold}.
A Riemann manifold is a Berwald–Finsler manifold, and not vice versa.
For Berwald–Finsler manifolds, the $hh-$curvature tensors reduce to
\begin{equation}
    \tensor{R}{_j^i_k_l}=\frac{\partial\Gamma^i_{jl}}{\partial x^k}-\frac{\partial\Gamma^i_{jk}}{\partial x^l}+\Gamma^i_{km}\Gamma^m_{jl}-\Gamma^i_{lm}\Gamma^m_{jk},
\end{equation}
similar to Riemann manifolds.

The \emph{flag curvature tensor} is defined as
\begin{equation}
    R_{ij}:=g_{ki}\tensor{R}{^k_j}=g_{ki}y^p\tensor{R}{_p^k_j_q}y^q=y^p\tensor{R}{_p_i_j_q}y^q=-\tensor{R}{_p_i_q_j}y^py^q.
\end{equation}
The \emph{Ricci scalar} is defined as
\begin{equation}
    Ric:=g^{ij}R_{ij}=\tensor{R}{^i_i}.
\end{equation}
The \emph{Ricci tensor} is defined as
\begin{equation}
    Ric_{ij}:=\frac{1}{2}(Ric)_{y^iy^j}.
\end{equation}
If $F$ is Riemann metric and $F^2=g_{ij}(x)y^iy^j$, there are relations such that
\begin{equation}
    \tensor{R}{^i_k}=\tensor{R}{_j^i_k_l}(x)y^jy^l, Ric=Ric_{ij}(x)y^iy^j,
\end{equation}
and $Ric_{ij}(x)$ is the Ricci tensor in Riemann manifolds.

\section{Modified Dispersion Relations and Finsler Geometry}\label{sec:mdrF}

A common feature among Lorentz violating theories is that most of them admit modified versions of dispersion relations for single particles.
When the energies of particles are far below the Planck scale, the MDRs for elementary particles can be written in the form of Taylor series as
\begin{equation}
    \begin{aligned}
E^2 & =m^2+p^2+D(p, \mu, M) \\
& =m^2+p^2+\sum_{n=1}^{\infty} \alpha_n(\mu, M) p^n,
\end{aligned}
\end{equation}
where $\alpha_n$ are dimensional coefficients, $\mu$ is some particle physics mass scale, and $M$ is the scale
associated with the new physics responsible for the correction of the dispersion relation, and usually taken to be the Planck scale $M_P\approx 10^{19}~\mathrm{GeV}.$
The spacetime structure corresponding to an MDR has been studied in the existent research (see, e.g.,~\cite{Kimberly2004}).
The key feature of the corresponding spacetime is described by a rainbow metric~\cite{rainbow}, which depends on the energy of the particle.
This is a natural concept from the QG point of view and arose in different contexts such as spacetime foam~\cite{Ellis2004},
the renormalization group applied to gravity~\cite{Girelli2007},  or as a consequence of averaging over QG fluctuations~\cite{Aloisio2006}.
However, the rainbow metric still lacks a rigorous formulation since it involves a metric defined on the tangent bundle while depending on a quantity associated with the cotangent bundle (the 4-momentum of the particle).
On the road to geometrical structures leading to MDRs, Riemannian geometry is abandoned, and the generalization of Riemannian geometry is adopted~\cite{Girelli2007}. This is Finsler geometry.

\subsection{The Bridge between Modified Dispersion Relations and Finsler Geometry}\label{sec:cons-mdrF}

The connection between MDRs and Finsler geometry was first proposed by Girelli  {et~al.}~\cite{Girelli2007-mdr}.
Here, we review the procedure in Ref.~\cite{Girelli2007-mdr}.

Starting from an MDR as $\mathcal{M}(p)=m^2$, one can suppose the action of a particle to be
\begin{equation}\label{eq:action}
    I=\int\left[\dot{x}^\mu p_\mu-\lambda\left(\mathcal{M}(p)-m^2\right)\right] d \tau,
\end{equation}
where $\lambda$ is a Lagrange multiplier that transforms appropriately under an arbitrary change of time parameter to ensure reparametrization invariance of the action.
Applying Hamilton equations on Equation~(\ref{eq:action}), one obtains
\begin{equation}
    p_\mu=\lambda \frac{\partial \mathcal{M}}{\partial\dot{x}^\mu}.
\end{equation}
If the relation above is invertible, one can rewrite the action in terms of velocities and the multiplier, hence obtaining
\begin{equation}
    I=\int {L}(x, \dot{x}, \lambda) d \tau.
\end{equation}
The multiplier can be eliminated by varying the action concerning it, and thus one obtains an action depending on $x$ and $\dot{x}$ as
\begin{equation}
    I=\int {L}(x, \dot{x}, \lambda(x, \dot{x})) d \tau.
\end{equation}
For a particle in Finsler spacetime, its action can be expressed as
\begin{equation}
    I=m\int F(x,\dot{x}) d\tau,
\end{equation}
so the Finsler norm is finally identified by
\begin{equation}\label{eq:finsler-lagrangian}
    m F(x,\dot{x})={L}(x, \dot{x}, \lambda(x, \dot{x})).
\end{equation}

\textls[-15]{To make the procedure clear, here we present the example~\cite{dsrFinsler} of the MDR for \mbox{DSR1~\cite{DSR1-1,DSR1-2}}} in a ``1  {+} 
 1'' spacetime.
The MDR for DSR1 can be expressed as
\begin{equation}\label{eq:mdr-dsr1}
    p_0^2-p_1^2-\ell p_0p_1^2=m^2,
\end{equation}
where $\ell$ is the LV scale.
The action takes the form
\begin{equation}
    I=\int\left(\dot{x}^\mu p_\mu-\lambda\left(p_0^2-p_1^2-\ell p_0 p_1^2-m^2\right)\right) d \tau,
\end{equation}
and the corresponding Hamilton equations are
\begin{equation}
\begin{aligned}
\dot{x}^0&=\lambda\left(2 p_0-\ell p_1^2\right), \\
\dot{x}^1&=\lambda\left(-2 p_1-2 \ell p_0 p_1\right) .
\end{aligned}
\end{equation}
Working at the leading order of $\ell$, inverting the above equation, one obtains
\begin{equation}
\begin{aligned}
p_0 & =\frac{\dot{x}^0}{2 \lambda}+\frac{\ell}{2} \frac{\left(\dot{x}^1\right)^2}{4 \lambda^2} \\
p_1 & =\frac{\dot{x}^1}{-2 \lambda}+\ell \frac{\dot{x}^0 \dot{x}^1}{4 \lambda^2};
\end{aligned}
\end{equation}
thus, the Lagrangian can be expressed at the leading order of $\ell$ as
\begin{equation}
    L=\frac{\left(\dot{x}^0\right)^2-\left(\dot{x}^1\right)^2}{4 \lambda}+\lambda m^2+\ell \frac{\dot{x}^0\left(\dot{x}^1\right)^2}{8 \lambda^2}.
\end{equation}
Minimizing the above equation with respect to $\lambda$, one obtains
\begin{equation}
    \lambda(\dot{x})=\frac{1}{2} \frac{\sqrt{\left(\dot{x}^0\right)^2-\left(\dot{x}^1\right)^2}}{m}+\frac{\ell}{2} \frac{\dot{x}^0\left(\dot{x}^1\right)^2}{\left(\dot{x}^0\right)^2-\left(\dot{x}^1\right)^2} .
\end{equation}
Thus, the momentum can be expressed by velocities as
\begin{equation}
\begin{aligned}
&p_0=\frac{m \dot{x}^0}{\sqrt{\left(\dot{x}^0\right)^2-\left(\dot{x}^1\right)^2}}-m^2 \frac{\ell}{2} \frac{\left(\dot{x}^1\right)^2\left(\left(\dot{x}^0\right)^2+\left(\dot{x}^1\right)^2\right)}{\left(\left(\dot{x}^0\right)^2-\left(\dot{x}^1\right)^2\right)^2},\\
&p_1=-\frac{m \dot{x}^1}{\sqrt{\left(\dot{x}^0\right)^2-\left(\dot{x}^1\right)^2}}+m^2 \ell \frac{\dot{x}^1\left(\dot{x}^0\right)^3}{\left(\left(\dot{x}^0\right)^2-\left(\dot{x}^1\right)^2\right)^2},
\end{aligned}
\end{equation}
and the Lagrangian is
\begin{equation}
    L=m\left(\sqrt{\left(\dot{x}^0\right)^2-\left(\dot{x}^1\right)^2}+\frac{\ell}{2} m \frac{\dot{x}^0\left(\dot{x}^1\right)^2}{\left(\dot{x}^0\right)^2-\left(\dot{x}^1\right)^2}\right).
\end{equation}
Finally, we obtain the Finsler norm associated with the MDR of DSR1 at the leading order of $\ell$ as
\begin{equation}\label{eq:finsler-dsr1}
    F(x,y)=\sqrt{\left(y^0\right)^2-\left(y^1\right)^2}+\frac{\ell}{2} m \frac{y^0\left(y^1\right)^2}{\left(y^0\right)^2-\left(y^1\right)^2},
\end{equation}
where $y^\mu=\dot{x}^\mu$ are the velocities.
For more examples, see Refs.~\cite{Girelli2007-mdr, Lobo2017-1, Lobo2017-2, Lobo2021, JieZ2022}.

In Ref.~\cite{Lobo2021}, Lobo and Pfeifer found a general formula connecting Finsler norms and general forms of MDRs in curled spacetime.
For an MDR described as
\begin{equation}
    m^2=g^{\mu\nu}(x)p_\mu p_\nu + \epsilon h(x,p),
\end{equation}
where $\epsilon$ is a perturbation parameter counting the first nontrivial contribution to the MDR, 
the corresponding Finsler norm can be expressed at the leading order of $\epsilon$ as
\begin{equation}\label{eq:finsler-general}
    F(x,\dot{x})=\sqrt{g_{\mu\nu}(x)\dot{x}^\mu\dot{x}^\nu}\left[1-\epsilon \frac{h(x,\bar{p}(x,\dot{x}))}{2m^2} \right],
\end{equation}
where 
\begin{equation}
    \bar{p}_\mu(x,\dot{x})=m \frac{\dot{x}_\mu}{\sqrt{g_{\mu\nu}(x)\dot{x}^\mu\dot{x}^\nu}}=m \frac{g_{\mu\nu}(x)\dot{x}^\nu}{\sqrt{g_{ab}(x)\dot{x}^a\dot{x}^a}}.
\end{equation}
For example, for an $n$th order  {polynomial modification} 
\begin{equation}
\begin{gathered}
{h(x, p)=h^{a_1 a_2 \ldots a_n}(x) p_{a_1} p_{a_2} \ldots p_{a_n},}\\
{\Rightarrow h(x, \bar{p}(x, \dot{x}))=m^n \frac{h_{a_1 a_2 \ldots a_n}(x) \dot{x}^{a_1} \dot{x}^{a_2} \ldots \dot{x}^{a_n}}{\left(g_{\mu\nu}(x)\dot{x}^\mu\dot{x}^\nu\right)^{\frac{n}{2}}},}
\end{gathered}
\end{equation}
and the Finsler norm is
\begin{equation}\label{eq:finsler-polynomial}
    F(x, \dot{x})=\sqrt{g_{\mu\nu}(x)\dot{x}^\mu\dot{x}^\nu}-\epsilon m^{n-2} \frac{h_{a_1 a_2 \ldots a_n}(x) \dot{x}^{a_1} \dot{x}^{a_2} \ldots \dot{x}^{a_n}}{2 \left(g_{\mu\nu}(x)\dot{x}^\mu\dot{x}^\nu\right)^{\frac{n-1}{2}}}.
\end{equation}

From Equation~(\ref{eq:finsler-lagrangian}), one can obtain the relation
\begin{equation}
    p_\mu=m \frac{\partial F}{\partial \dot{x}^\mu}=m \frac{g_{\mu \nu} \dot{x}^\nu}{F},
\end{equation}
where $g_{\mu\nu}$ is the Finsler metric of $F(x,\dot{x})$ defined in Equation~(\ref{eq:gij}).
Inverting the above equation, and using the inverse metric $g^{\mu\nu}$, one can recover the MDR $\mathcal{M}(p)=m^2$ as
\begin{equation}
    m^2=g^{\mu \nu}(\dot{x}(p)) p_\mu p_\nu.
\end{equation}
From Equation~(\ref{eq:action}) one can obtain
\begin{equation}
    \dot{x}^\mu=\lambda \frac{\partial \mathcal{M}(p)}{\partial p_\mu},
\end{equation}
thus the traditional relation
\begin{equation}
    v_i=\frac{\dot{x}^i}{\dot{x}^0}=\frac{\partial p_0}{\partial p_i} 
\end{equation}
holds.

From Equations~(\ref{eq:finsler-dsr1}) and~(\ref{eq:finsler-polynomial}), 
we can see that the mass of the particle $m$ plays an important role in the Finsler norm.
Keeping the MDR unchanged and letting $m\rightarrow 0$, we can see that the above Finsler norm becomes a trivial Riemann norm.
Thus, the procedure fails when meeting MDRs with zero mass, such as photons.
In addition, since the Finsler norm is an order-1 homogeneous function of $\dot{x}$, any other parameters with mass dimensions in the Finsler norm should be present as combinations to cancel their dimensions.
To insert the LV scale into the Finsler norm, we need other parameters to cancel the dimension introduced by the LV scale.
Since we construct the Finsler norm from an MDR and without any assumption introducing dimensional physical constants, 
the only parameter to cancel the dimension from the LV scale is the mass of the particle.
For example, in the MDR as Equation~(\ref{eq:mdr-dsr1}), the dimension of the LV scale $\ell$ is finally canceled in the Finsler norm by $m$ as a combination of $m\ell$ in Equation~(\ref{eq:finsler-dsr1}). 
Therefore, it is not feasible to construct Finsler norms from MDRs for massless particles without other assumptions.
Another feature we should notice is that even with the same MDR, different particles with different masses meet different Finsler structures.

\subsection{Physical Influences from Finsler Geometry}\label{sec:physicsF}

\subsubsection{Time Dilation in Finsler Geometry}
Time dilation is an important phenomenon in special relativity.
It enlarges the lifetime of high-energy particles in high-energy particle physics experiments.
In Finsler geometry, the length element is modified and the time dilation formula is modified.
Lobo and Pfeifer realized this phenomenon~\cite{Lobo2021, Lobo2022}, and here we review their results.

In Finsler spacetime, the proper time of a massive particle experiencing between events $A$ and $B$ along a time-like curve can be defined as
\begin{equation}
    \Delta \tau_{AB}=\int_A^B F(x^\mu,\dot{x}^\mu)d\tau.
\end{equation}
Noticing that $x^0=t$ is the time coordinate of the local observer, the time dilation can be expressed by
\begin{equation}
    \Delta \tau =  F(x^\mu, \frac{d x^\mu}{d x^0})\cdot \Delta t,
\end{equation}
where $\Delta \tau$ is the proper time duration of the particle and $\Delta t$ is the observed time duration.
Taking the MDR of DSR1 as an example, the Finsler norm is expressed as Equation~(\ref{eq:finsler-dsr1}); thus, the time dilation can be expressed as
\begin{equation}
    \Delta \tau=\frac{\Delta t}{\gamma}\left[1+\frac{\ell}{2} m \gamma\left(\gamma^2-1\right)\right],
\end{equation}
where $\gamma$ is the usual velocity Lorentz factor as
\begin{equation}
    \gamma:=\frac{1}{\sqrt{1-v^2}},
\end{equation}
with $v^i:=dx^i/dx^0$ and $v^2=\delta_{ij}v^iv^j$.
At the leading order of $\ell$, the above time dilation can be expressed as
\begin{equation}\label{eq:td}
    \Delta t=\gamma \left[1-\frac{\ell}{2} m \gamma\left(\gamma^2-1\right)\right]\cdot \Delta \tau.
\end{equation}
Considering that
\begin{equation}\label{eq:pvdsr1}
\begin{aligned}
& p_0=m \frac{\partial}{\partial \dot{x}^0} F(x, \dot{x})=m \gamma-\frac{\ell}{2} m^2\left(\gamma^2-1\right)\left(2 \gamma^2-1\right),\\
& p_i=m \frac{\partial}{\partial \dot{x}^i} F(x, \dot{x})=-v_i \gamma m+\ell m^2 v_i \gamma^4,
\end{aligned}
\end{equation}
we can express the Lorentz factor with the energy of the particle at the leading order of $\ell$ as
\begin{equation}
    \gamma=\frac{p_0}{m}+\frac{\ell}{2} m\left(1-3 \frac{p_0^2}{m^2}+2 \frac{p_0^4}{m^4}\right).
\end{equation}
In the limit of $p_0\gg m$, Equation~(\ref{eq:td}) can be expressed as
\begin{equation}
\begin{aligned}
 \Delta t&=\frac{p_0}{m} \Delta \tau\left[1+\frac{\ell}{2}\left(\frac{m^2}{p_0}-2 p_0+\frac{p_0^3}{m^2}\right)\right]\\
 &\approx \frac{p_0}{m} \Delta \tau\left[1+\frac{\ell}{2}\frac{p_0^3}{m^2}\right]\\
 &=\Delta t_{\rm SR} \left[1+\frac{\ell}{2}\frac{p_0^3}{m^2}\right],   
\end{aligned}
\end{equation}
where $\Delta t_{\rm SR}=\frac{p_0}{m} \Delta \tau$ is the observed time in special relativity.
Thus, the limit on $\ell$ of unstable particles can be made from experiments, i.e., muons~\cite{Lobo2021}.

\subsubsection{Arrival Time Delay of Astroparticles in Finsler Geometry}

One important influence on particles from LV is the possible speed variation.
By the relation $v=\partial E/\partial p$, the velocities of particles are modified from the modification of dispersion relations, 
especially when the particles have high energies, such as astroparticles.
After propagating cosmological distances, the astroparticles from the same source emitted at the same time will arrive at the Earth's equipment at different times because of the speed variation.
This effect opens a window for constraining the LV scale. 
Tests on LV based on the arrival time delay of astroparticles can be found in Refs.~\cite{Ellis2003, Amelino2003, Ellis2006, Jacob2007, Biesiada2009, Shao2010-1, Shao2010-2, fermi2013, Zhang2015, Pan2015, Xu2016-1, Xu2016-2, Zou2018, Liu2018, Xu2018, Huang2018, Huang2019, Li2020, Pan2020, Acciari2020, Chen2021, JieZ2021-1, JieZ2021-2}.

Since the Universe is not flat, there arises the question as to how to calculate the arrival time differences between different astroparticles with different energies under the influence of LV.
Jacob and Piran~\cite{Jacob2008} derived the time delay formula in the standard model of cosmology.
For an MDR expressed as
\begin{equation}\label{eq:mdrtd}
    E^2=m^2+p^2\left[1-s_n\left(\frac{p}{E_{\mathrm{LV}, n}}\right)^n\right]=m^2+p^2+\alpha p^{n+2},
\end{equation}
the arrival time delay between massless particles with high energies and low energies (which is assumed negligible) is
\begin{equation}\label{eq:tdPrian}
    \Delta t=\frac{1+n}{2 H_0}\left(\frac{E_{\mathrm{obs}}}{E_{\mathrm{LV}, n}}\right)^n \int_0^z \frac{\left(1+z^{\prime}\right)^n \mathrm{~d} z^{\prime}}{\sqrt{\Omega_{\mathrm{m}}\left(1+z^{\prime}\right)^3+\Omega_{\Lambda}}},
\end{equation}
where $z$ is the redshift of the source of the two particles, $E_{\text {obs }}$ is the observed energy of the high-energy particle from Earth  {equipment, } 
$\Omega_{\mathrm{m}}$ and $\Omega_{\Lambda}$ are universe constants, and $H_0$ is the current Hubble parameter. 
Generalized research on the time delay from homogeneous and isotropic modified dispersion relations in a Friedmann–Lemaître–Robertson–Walker (FRW) universe is presented in Ref.~\cite{Pfeifer2018}. 
In this work, the MDR in the FRW universe is assumed to be
\begin{equation}
    H\left(t, p_t, w\right)=-p_t^2+a(t)^{-2} w^2+\epsilon h\left(t, p_t, w\right),
\end{equation}
where 
\begin{equation}
    w^2=p_r^2 \chi^2+\frac{p_\theta^2}{r^2}+\frac{p_\phi^2}{r^2 \sin ^2 \theta}, \quad \chi=\sqrt{1-k r^2},
\end{equation}
$h\left(t, p_t, w\right)$ can be an arbitrary function of $t, p_t $, and $w$, and $\epsilon$ is an arbitrary perturbation parameter.
The time delay formula can be obtained by solving the Hamilton equations of motion of the above Hamiltonian.
Now that the Finsler spacetimes corresponding to MDRs are constructed, it is time to reconsider the time delay problem via the geometrical approach by solving geodesic equations.

In Section~\ref{sec:cons-mdrF} the Finsler spacetimes corresponding to MDRs were constructed. However, for MDRs in flat spacetime, the constructed Finsler norm is also flat. It is not clear how to corporate the additional LV term in the Finsler metric with a background Riemann metric.
In Refs.~\cite{Lobo2017-1, Lobo2017-2}, Lobo  {et~al.}~bypassed the difficulty by assuming the form of the MDR in the curved spacetime at the beginning.
In Ref.~\cite{Lobo2017-1}, they assumed that the MDR which formalizes the physics of particles embedded in a de Sitter-like curved spacetime can be expressed as
\begin{equation}
    m^2=p_0^2-p_1^2 e^{-2 H x^0}+\ell\left(\gamma p_0^3+\beta p_0 p_1^2 e^{-2 H x^0}\right),
\end{equation}
where $H$ is the parameter of curvature, $\ell$ is the deformation parameter due to Planck-scale effects, and $\beta, \gamma$ are two numerical parameters of order 1.
In Ref.~\cite{Lobo2017-2}, they focused on the MDR as
\begin{equation}
    m^2=a^{-2}(\eta)\left(\Omega^2-\Pi^2\right)+\ell a^{-3}(\eta)\left(\gamma \Omega^3+\beta \Omega \Pi^2\right),
\end{equation}
where $(\eta,x)$ are the so-called conformal time coordinates, $(\Omega, \Pi)$ are their conjugate momenta, $a(\eta)$ is the scale factor of the universe, $\ell$ is the deformation parameter due to Planck-scale effects, and $\beta, \gamma$ are two numerical parameters of order 1.
By constructing the corresponding Finsler spacetime and solving the geodesic equations, they obtained the time delay formula.

Luckily, the FRW metric for the real universe is simple and can be expressed as
\begin{equation}
    ds^2=dt^2-a(t)^2(dx^2+dy^2+dz^2), 
\end{equation}
or written in Finsler norm as
\begin{equation}
    F_{\rm FRW}=\sqrt{(y^0)^2-(a(x^0)y^1)^2-(a(x^0)y^2)^2-(a(x^0)y^3)^2}.
\end{equation}
It is natural to embed the effect of the expansion of the universe into the Finsler norm just by replacing $y^i$ with $a(x^0)y^i$, where $i$ is the space index.
It makes sense since $a(x^0)$ describes how the space expands and it should be multiplied to every space component in the metric.
In Ref.~\cite{JieZ2022}, the above strategy was applied. The Finsler norm in FRW universe corresponding to the MDR in Equation~(\ref{eq:mdrtd}) in ``1  {+} 1'' spacetime is
\begin{equation}
F= \sqrt{\left(y^0\right)^2-\left(a\left(x^0\right) y^1\right)^2} +\alpha m^n \frac{\left(-a\left(x^0\right) y^1\right)^{n+2}}{2\left(\sqrt{\left(y^0\right)^2-\left(a\left(x^0\right) y^1\right)^2}\right)^{n+1}},
\end{equation}
and the geodesic equations are
\begin{adjustwidth}{-\extralength}{-2cm}
\begin{equation}
\begin{gathered}
{\dot{y}^0+a\left(x^0\right) a^{\prime}\left(x^0\right)\left(y^1\right)^2+
\alpha m^n \frac{(n+2) a^{\prime}\left(x^0\right) a\left(x^0\right)^{n+1}\left(-y^1\right)^{n+2}\left[(n-1)\left(y^0\right)^2+a\left(x^0\right)^2\left(y^1\right)^2\right]}{2\left[\left(y^0\right)^2-a\left(x^0\right)^2\left(y^1\right)^2\right]^{\frac{n+2}{2}}}=0,}\\
{\dot{y}^1+2 \frac{a^{\prime}\left(x^0\right)}{a\left(x^0\right)} y^0 y^1-\alpha m^n \frac{n(n+2) a^{\prime}\left(x^0\right) a\left(x^0\right)^{n-1}\left(y^0\right)^3\left(-y^1\right)^{n+1}}{2\left[\left(y^0\right)^2-a\left(x^0\right)^2\left(y^1\right)^2\right]^{\frac{n+2}{2}}}=0 .}
\end{gathered}
\end{equation}
\end{adjustwidth}
The solution to the above geodesic equations suggests the same time delay formula as Equation~(\ref{eq:tdPrian}), obtained by Jacob and Piran differently from the standard model of cosmology. 

One thing that should be noticed is that the Finsler norms corresponding to MDRs strongly rely on the forms of dispersion relations. 
Even if two forms of dispersion relations break Lorentz invariance in the same order, from Equation~(\ref{eq:finsler-polynomial}) we can see that the corresponding Finsler norms can be different. 
Thus, the geodesic equations are different.
However, interestingly, the differences do not influence the arrival time delay in Finsler spacetimes when the modified term of the MDR is not zero in the limit $p/E\rightarrow 1$~\cite{JieZ-unp}.
For the MDR of the form
\begin{equation}
    E^2=m^2+p^2+\alpha E^{n+2} G\left(\frac{p}{E}\right),
\end{equation}
the arrival time delay is calculated as
\begin{equation}
    \Delta t=\alpha G(1) E_\mathrm{obs}^n \frac{n+1}{2H_0}\int_0^z \frac{(1+z')^n}{\sqrt{\Omega_{m}\left(1+z^{\prime}\right)^{3}+\Omega_{\Lambda}}}dz'.
\end{equation}
Thus, if $G(1)\ne 0$, we can define new $\alpha^\prime= \alpha G(1)$ and $G^\prime(x)=G(x)/G(1)$, and the above time delay formula is still the same as Equation~(\ref{eq:tdPrian}). 
If $G(1)=0$, which means the modified term in the MDR becomes zero in the limit $p/E\rightarrow 1$, there is no time delay.
An example for the $G(1)=0$ is the dispersion relation of DSR2~\cite{DSR2-1,DSR2-2}, expressed as
\begin{equation}
    \frac{E^2-p^2}{(1-\lambda E)^2}=\frac{m^2}{(1-\lambda m)^2}=\mu^2
\end{equation}
or $E^2=\mu^2+p^2-2\lambda E(E^2-p^2)$ at the leading order of $\lambda$.

The derivation of the time delay formula brings success to describing LV with Finsler geometry.
However, determining how to embed a background Riemann metric into a flat Finsler norm describing LV is still an open question.
This question arises when studying a particle subject to LV moving in curled spacetimes, e.g., black holes or gravity lensing.

\subsubsection{Transformation between Inertial Frames and Modified Composition Laws}

The principle of relativity plays an important role in Einstein's general relativity. 
It requires that the equations describing the laws of physics have the same form in all admissible frames of reference.
In Section~\ref{sec:cons-mdrF}, we construct the Finsler spacetime describing an MDR, and the formulas of many physical quantities are modified, such as the relations between the momentum $p_\mu$ and the speed $v$.
This causes a serious problem: if the composition law of the momentum is still the traditional adding law such as $p_{1\mu}+p_{2\mu}=p_{3\mu}+p_{3\mu}$ in one inertial frame, after frame transformation, the equality can not be expressed as $p_{1\mu}^\prime+p_{2\mu}^\prime=p_{3\mu}^\prime+p_{3\mu}^\prime$, where $p_{i\mu}^\prime$ is the new 4-momentum in the new frame.
Thus, a new composition law of momentum is needed in the Finsler spacetime.
A natural idea is an assumption that the principle of reality still holds in Finsler spacetime.
Here, we review the results of Ref.~\cite{Lobo2022}.

To start the discussion, here, we take the MDR of DSR1 in (1+1) spacetime as an example.
The MDR can be expressed as
\begin{equation}\label{eq:mdr-dsr1new}
    m^2=p_0^2-p_1^2-\ell p_0p_1^2,
\end{equation}
the Finsler norm is Equation~(\ref{eq:finsler-dsr1}), and the relations between the momentum $p_\mu$ and the speed $v$ are Equation~(\ref{eq:pvdsr1}).
Next, let us consider the transformation law in two inertial frames.
Let us assume that the two inertial frames are labeled by $S$ and $\tilde{S}$ and move with relative velocity $v$, and the observers in $S$ and $\tilde{S}$ assign momentum $p_\mu$ and $\Tilde{p}_\mu$ for a particle.
From Equation~(\ref{eq:pvdsr1}), at the leading order of $\ell$ the relation between $p_\mu$ and $\Tilde{p}_\mu$ can be assumed to be
\begin{equation}
\begin{aligned}
& \tilde{p}_0=\gamma\left(p_0-v p_1\right)+\ell\left[A p_0 p_1+B p_1^2-\frac{1}{2} p_0^2\left(\gamma^2-1\right)\left(2 \gamma^2-1\right)\right], \\
& \tilde{p}_1=\gamma\left(p_1-v p_0\right)+\ell\left(p_0^2 v \gamma^4+F p_0 p_1+G p_1^2\right),
\end{aligned}
\end{equation}
where $\gamma=\frac{1}{\sqrt{1-v^2}}$, and $A, B, F$, and $G$ are functions of $v$.
The principle of relativity commands that $m^2=\Tilde{p}_0^2-\Tilde{p}_1^2-\ell \Tilde{p}_0\Tilde{p}_1^2$.
Thus, we have the relation 
\begin{equation}
\begin{aligned}
& B=-\frac{A}{v}-\frac{\left(1-v^2\right)^{3 / 2}-v^2-1}{2\left(1-v^2\right)}, \\
& F=-\frac{A}{v}, \\
& G=A-\frac{v\left[2 v^2-\left(1-v^2\right)^{3 / 2}\right]}{2\left(1-v^2\right)^2}.
\end{aligned}
\end{equation}
In Ref.~\cite{Lobo2022} these parameters are fixed by setting $A=0$ for simplicity.
The reader should keep in mind that more assumptions are needed to fix all of the parameters.
From the above relations and setting $A=0$, the transformation relations are
\begin{adjustwidth}{-\extralength}{0cm}
\begin{equation}
[\Lambda(v, p)]_\mu=\tilde{p}_\mu=\left\{\begin{array}{l}
\tilde{p}_0=\gamma\left(p_0-v p_1\right)+\frac{\ell}{2}\left[p_1^2 \gamma\left(2 \gamma^3-\gamma-1\right)-p_0^2\left(\gamma^2-1\right)\left(2 \gamma^2-1\right)\right], \\
\tilde{p}_1=\gamma\left(p_1-v p_0\right)+\ell v\left[p_0^2 \gamma^4-\frac{p_1^2}{2} \gamma\left(2 \gamma^3-2 \gamma-1\right)\right],
\end{array}\right.
\end{equation}
\end{adjustwidth}
where $[\Lambda(v, p)]_\mu$ is referred to be the transformed $\mu$-component of momenta $p$ using boost parameter $v$.

Now, we turn to the modified composition law. 
The most general form of the composition law in first-order perturbation is
\begin{equation}
\begin{aligned}
& (p \oplus q)_0=p_0+q_0+\ell\left(\alpha p_0 q_0+\beta p_1 q_1+\omega p_0 q_1+\eta p_1 q_0\right),\\
& (p \oplus q)_1=p_1+q_1+\ell\left(\delta p_1 q_0+\epsilon p_0 q_1+\lambda p_1 q_1+\mu p_0 q_0\right), 
\end{aligned}
\end{equation}
where $(\alpha, \beta,\omega,\eta,\delta,\epsilon,\lambda,\mu)$ are dimensionless parameters yet to be determined.
The momentum composition law and the translation relation must have a relation as
\begin{equation}
    \Lambda(v, p \oplus q)=\Lambda\left(v_q, p\right) \oplus \Lambda\left(v_p, q\right),
\end{equation}
where $v_p$ and $v_q$ are back-reacting parameters and may depend on the moment $p$ and $q$, respectively.
The most general relations on $v_p$ and $v_q$ are
\begin{equation}
\begin{aligned}
& v_q=v+\ell\left(H q_0+J q_1\right),\\
& v_p=v+\ell\left(M p_0+R p_1\right).
\end{aligned}
\end{equation}
The parity invariance demands that the composition law is invariant under transformations $p_0\rightarrow p_0$ and $ p_i\rightarrow -p_i$, and this demands $\omega=\eta=\mu=\lambda=0$.
Another thing that should be considered is the commutative property and the associative property of the composition law, which means that $p\oplus q=q\oplus p$ and $(p\oplus q)\oplus r=p\oplus(q\oplus r)$, and these conditions demand $\eta=\omega$ and $\delta=\epsilon$.
These constraints suggest that
\begin{equation}
\begin{gathered}
{\alpha=\omega=\eta=\mu=\lambda=0,\quad \epsilon=\delta, \quad \beta=1+2\delta,}\\
{R=J=\frac{1-\gamma}{\gamma}\left(1+\gamma-\frac{\delta}{\gamma}\right), \quad M=H=-v\left(\gamma+\frac{\delta}{\gamma}\right),}
\end{gathered}
\end{equation}
and the modified composition law
\begin{equation}\label{eq:mcl-delta}
\left\{\begin{array}{l}
(p \oplus q)_0=p_0+q_0+(1+2\delta)\ell p_1 q_1,\\
(p \oplus q)_1=p_1+q_1+\delta\ell(p_0 q_1+p_1 q_0). 
\end{array}\right.    
\end{equation}
Equation~(\ref{eq:mcl-delta}) suggests one more constraint to obtain the value of $\delta$.
The undeformed spatial momentum conservation requires $\delta=0$, and the  composition law becomes
\begin{equation}
\left\{\begin{array}{l}
(p \oplus q)_0=p_0+q_0+\ell p_1 q_1,\\
(p \oplus q)_1=p_1+q_1. 
\end{array}\right.    
\end{equation}
The undeformed energy conservation requires $\delta=-1/2$, and the composition law becomes
\begin{equation}
\left\{\begin{array}{l}
(p \oplus q)_0=p_0+q_0,\\
(p \oplus q)_1=p_1+q_1-\frac{1}{2}\ell(p_0 q_1+p_1 q_0). 
\end{array}\right.    
\end{equation}
Another case is $\delta=-1$. In this case, the composition law becomes
\begin{equation}
\left\{\begin{array}{l}
(p \oplus q)_0=p_0+q_0-\ell p_1 q_1,\\
(p \oplus q)_1=p_1+q_1-\ell(p_0 q_1+p_1 q_0). 
\end{array}\right.   
\end{equation}
This composition law is the one suggested in DSR1 from Refs.~\cite{DSR1-2, Lukierski2003, Judes2003}.
In Ref.~\cite{Lobo2022}, one kind of conservation law without the commutative proper was also considered.
This kind of conservation law reads
\begin{equation}
\left\{\begin{array}{l}
(p \oplus q)_0=p_0+q_0,\\
(p \oplus q)_1=p_1+q_1-\ell p_0 q_1. 
\end{array}\right.  
\end{equation}
This composition law is the one from the bicrossproduct $\kappa$-Poincaré coproduct structure~\cite{Gubitosi2013}.

\section{Connections Between Lorentz Violation Theories and Finsler Geometry} \label{sec:lvF}

\subsection{Doubly Special Relativity and Finsler Geometry}\label{sec:DSR}

Doubly special relativity is a novel idea about the fate of Lorentz invariance.
From a phenomenological perspective, DSR is a set of assumptions that the Lorentz group acts on so that the usual speed of light, c, and a new momentum scale, $E_{\rm DSR}$, are invariant.
Usually, $E_{\rm DSR}$ is taken as the Planck energy.
Soon after the idea was pointed out by Amelino-Camelia~\cite{DSR1-1, DSR1-2},
$\kappa$-Poincaré algebra~\cite{Lukierski1991-1, Lukierski1991-2, Majid1994, Lukierski1995} was applied on this kind of DSR~\cite{Kowalski2001, Bruno2001, Kowalski2002, Lukierski2003}.
This kind of DSR was referred to as DSR1 later.
In DSR1, the brackets of rotations $M_i$ boosts $N_i$, and the components of momenta $P_\mu$ read
\begin{equation}
\begin{gathered}
{\left[M_i, M_j\right]=i \epsilon_{i j k} M_k, \quad\left[M_i, N_j\right]=i \epsilon_{i j k} N_k,\quad {\left[N_i, N_j\right]=-i \epsilon_{i j k} M_k,}} \\
{\left[M_i, P_j\right]=i \epsilon_{i j k} P_k, \quad\left[M_i, P_0\right]=0,} \\
{\left[N_i, P_j\right]=i \delta_{i j}\left(\frac{\kappa}{2}\left(1-e^{-2 P_0 / \kappa}\right)+\frac{1}{2 \kappa} \vec{P}^2\right)-i \frac{1}{\kappa} P_i P_j,} \\
{\left[N_i, P_0\right]=i P_i,}
\end{gathered}
\end{equation}
where $\kappa$ acts as the role of $E_{\rm DSR}$.
The Casimir of the $\kappa$-Poincaré algebra reads
\begin{equation}\label{eq:casimir-dsr1}
    \kappa^2 \cosh \frac{P_0}{\kappa}-\frac{\vec{P}^2}{2} e^{P_0 / \kappa}=M^2.
\end{equation}
Expanding it at the leading order of $\frac{1}{\kappa}$, one obtains
\begin{equation}\label{eq:casimir-dsr1m}
    P_0^2-\vec{P}^2-\ell P_0 \vec{P}^2=m^2,
\end{equation}
where $\ell=1/\kappa$ and $m^2=2(M^2-\kappa^2)$.
From Equation~(\ref{eq:casimir-dsr1}), one can obtain the result that the value of three-momentum $|\Vec{P}|=\kappa$ corresponds to infinite energy $P_0=\infty$. Thus, in DSR1 there is a maximal three-momentum $|\Vec{P}|=\kappa$, and if a particle has momentum $|\Vec{P}|=\kappa$ for some observers, it has the same momentum for all observers.

Another realization of DSR, called DSR2, was proposed by Magueijo and Smolin~\cite{DSR2-1, DSR2-2}.
In DSR2, the Lorentz algebra is not deformed, and the boosts–momenta generators now have the form
\begin{equation}
    \left[N_i, P_j\right]=i\left(\delta_{i j} P_0-\lambda P_i P_j\right), \quad \left[N_i, P_0\right]=i\left(1-\lambda{P_0}\right) P_i,
\end{equation}
where $\lambda=\frac{1}{E_{\rm DSR}}$.
The Casimir of the above algebra reads
\begin{equation}\label{eq:casimir-dsr2}
    M^2=\frac{P_0^2-\Vec{P}^2}{(1-\lambda P_0)^2}.
\end{equation}
From Equation~(\ref{eq:casimir-dsr2}), one obtains that $P_0=1/\lambda$ corresponds to infinite momentum, and this energy $1/\lambda$ is invariant in transformations of frames.

Although the DSR theories are referred to as Lorentz-violation theories, the symmetry group is still the Lorentz group, but acting nonlinearly on the energy-momentum \mbox{sector~\cite{DSR2-1, Kowalski2002, Lukierski2003}.}
The nonlinear representations lead to modified conservation laws.
In Ref.~\cite{Judes2003}, Judes proposed general conservation laws for DSRs.
Since a DSR symmetry group is simply a nonlinear realization of the Lorentz group,
a function between the physical energy–momentum $P_4=(E, P)$ and a pseudo-energy–momentum $\mathcal{P}_4=(\epsilon,\pi)$, which transforms similar to a Lorentz 4-vector, can be established as
\begin{equation}
    P_4=F(\mathcal{P}_4),\quad \mathcal{P}_4=F^{-1}(P_4).
\end{equation}
The transformations act on the pseudo-energy–momentum in the normal linear manner $\mathcal{P}_4^\prime=\mathcal{L}(\mathcal{P}_4)$,
where $\mathcal{L}$ is the usual Lorentz transformation.
Function $F$ satisfies the dispersion relation
\begin{equation}
    \left[\epsilon(E,p)\right]^2-\left[\pi(E,p)\right]^2=\mu_0^2,
\end{equation}
where $\mu_0$ is the Casimir invariant and $\mu_0=\epsilon(m,0)$.
In the linear representation, the total energy can be defined as
\begin{equation}
    \mathcal{P}_4^{\rm t o t}=\sum_i \mathcal{P}_4^i;
\end{equation}
thus, in the nonlinear representation, the total energy can be expressed as
\begin{equation}
    P_4^{\rm t o t}=F\left(\sum_i F^{-1}\left(P_4^i\right)\right).
\end{equation}
As an example, in the case of DSR2, the relation between $(E,p)$ and $(\epsilon,\pi)$ can be expressed~as
\begin{equation}
    (\epsilon,\pi)=\frac{(E,p)}{1-\lambda E},\quad (E,p)=\frac{(\epsilon,\pi)}{1+\lambda \epsilon};
\end{equation}
thus, the conservation law reads
\begin{equation}
    \epsilon_{\rm t o t}=\sum_i \frac{E_i}{1-\lambda E_i}, \quad \pi_{\rm t o t}=\sum_i \frac{p_i}{1-\lambda E_i},
\end{equation}
and
\begin{equation}\label{eq:conservation}
    E_{\rm t o t}=\frac{\sum_i E_i /\left(1-\lambda E_i\right)}{1+\lambda \sum_i E_i /\left(1-\lambda E_i\right)}, \quad p_{\rm t o t}=\frac{\sum_i p_i /\left(1-\lambda E_i\right)}{1+\lambda \sum_i E_i /\left(1-\lambda E_i\right)} .
\end{equation}

The formulation of DSR in the energy–momentum space is clearly incomplete,
as it lacks any description of the structure of spacetime. 
DSR was formulated in a somehow unusual way: one started with the energy–momentum space and only then the problem of construction of spacetime was considered.
An early discussion on the connection between DSR and Finsler geometry can be found in Ref.~\cite{Mignemi2007}, via the relation between the MDRs of DSRs and Finsler geometries.
In Ref.~\cite{dsrFinsler}, Amelino-Camelia  {et~al.}~connected DSR1 symmetries with Finsler spacetime.
The Finsler norm for DSR1 is Equation~(\ref{eq:finsler-dsr1}); for convenience, here we write it as
\begin{equation}\label{eq:finsler-dsr1m}
    F(x,y)=\sqrt{\left(y^0\right)^2-\left(y^1\right)^2}+\frac{A}{2} \frac{y^0\left(y^1\right)^2}{\left(y^0\right)^2-\left(y^1\right)^2},
\end{equation}
where $A=m\ell$ is a dimensionless parameter.
Well-defined relationships between modified relativistic symmetries and Finsler geometries are established in their work.

However, one huge difference between DSR theories and Finsler geometries can be seen in Equation~(\ref{eq:finsler-dsr1m}).
In DSR theories, the new momentum scale $E_{\rm DSR}$ ($\frac{1}{\ell}$ in DSR1) is invariant under transformations in different frames.
This parameter is supposed to be a constant of the universe at the beginning of the construction of the theories.
However, from Equation~(\ref{eq:finsler-dsr1m}), the structure of the Finsler spacetime is described by the parameter 
\begin{equation}
    A=m\ell=\frac{m}{E_{\rm DSR}}.
\end{equation}
This means that if the new momentum scale $E_{\rm DSR}$ is indeed a constant of the universe, different particles with different masses meet different Finsler structures.
On the other hand, if one assumes that the universe has a unique Finsler structure, the energy scale $E_{\rm DSR}$ depends on the particle mass as $E_{\rm DSR}=\frac{m}{A}$ and is not a constant of the universe.

Another thing that needs to be mentioned is the problem when DSR is applied to macroscopic objects.
In fact, DSR is a nonperturbative theory.
In the macro world, the resting energies of objects can easily achieve $E_{\rm DSR}$, and this means obvious effects on the modified relations, e.g., the conservation laws shown as Equation~(\ref{eq:conservation}).
To solve this problem, Magueijo and Smolin~\cite{DSR2-2} proposed that $E_{\rm DSR}$ in a system of $N$ elementary particles should be replaced by $NE_{\rm DSR}$.
This proposal is just an assumption in their work, but now it has roots in Finsler geometry.
We consider a macro system with $N$ same particles moving with the same speed $v$ and assume that these particles have mass $m$, and they are of DSR1 particles with an MDR as Equation~(\ref{eq:casimir-dsr1m}).
Thus, they have 
 the same Finsler structure as Equation~(\ref{eq:finsler-dsr1m}), with the same constant $A$.
Now let us regard these particles as a whole.
Since they have 
 the same Finsler structure and have the same speed, the whole system should also be under the effect of the same Finsler structure.
This system has mass $Nm$ and Finsler constant $A$; thus, the momentum scale of the system becomes
\begin{equation}
    E=\frac{Nm}{A}=NE_{\rm DSR},
\end{equation}
which is the same as the proposal of Magueijo and Smolin.

\subsection{Standard-Model Extension and Finsler Geometry}\label{sec:SME}

The Standard-Model Extension~\cite{cpt-sm,lv-sm,glv-sm} is a comprehensive realistic effective field with Lorentz violation that incorporates both the Standard Model and general relativity.
The connection between SME and Finsler geometry was established in Refs.~\cite{finsler-sme,bipartite-sme,Colladay2015,Russell2015,Schreck2016,finsler-scalar,sme-higher-order1,sme-higher-order2}.  

For a single massive spin-$\frac{1}{2}$ Dirac fermion, the general form of the quadratic sector of a renormalizable Lorentz- and $CPT$-violating Lagrangian~\cite{cpt-sm,lv-sm} is
\begin{equation}\label{eq:smefermion}
    \mathcal{L}=\frac{1}{2} i \bar{\psi} \Gamma^\nu \stackrel{\leftrightarrow}{\partial}_\nu \psi-\bar{\psi} M \psi,
\end{equation}
where
\begin{equation}
    \Gamma^\nu:=\gamma^\nu+c^{\mu \nu} \gamma_\mu+d^{\mu \nu} \gamma_5 \gamma_\mu+e^\nu+i f^\nu \gamma_5+\frac{1}{2} g^{\lambda \mu \nu} \sigma_{\lambda \mu}
\end{equation}
and
\begin{equation}
    M:=m+a_\mu \gamma^\mu+b_\mu \gamma_5 \gamma^\mu+\frac{1}{2} H^{\mu \nu} \sigma_{\mu \nu}.
\end{equation}
To study the corresponding classical Lorentz-violating kinematics, a topic central to subjects such as the behavior of quantum wave packets, the analysis of relativistic scattering, and the motion of macroscopic bodies, one useful approach is to introduce an analog point-particle system with relativistic Lagrangian $L$, which leads directly to various results such as the equations of motion for the classical trajectory, the momentum–velocity connection, and the dispersion relation. Thus, Finsler geometry is connected with SME via the relativistic Lagrangian $L$ by
\begin{equation}
    L=-m F(x,y).
\end{equation}

The MDR of the Lagrangian Equation~(\ref{eq:smefermion}) was derived in Ref.~\cite{sclcpt} as
\begin{equation}
    \mathrm{det} \left(\Gamma^\mu p_\mu-M\right)=0.
\end{equation}
Expansion of the determinant of this matrix yields
\begin{equation}\label{eq:smemdr}
    \begin{aligned}
0= & \frac{1}{4}\left(V^2-S^2-A^2-P^2\right)^2+4[P(V T A)-S(V \widetilde{T} A)-V T T V+A T T A] \\
& +V^2 A^2-(V \cdot A)^2-X\left(V^2+S^2-A^2-P^2\right)-2 Y S P+X^2+Y^2,
\end{aligned}
\end{equation}
where the scalar quantity is $S=-m+e \cdot p$, 
the pseudoscalar is $P=f \cdot p$, 
the vector is $V_\mu=p_\mu+(c p)_\mu-a_\mu$, 
the axial-vector is $A_\mu=(d p)_\mu-b_\mu$, 
and the tensor is $T_{\mu \nu}=\frac{1}{2}(g p-H) \mu \nu$. 
The two invariants of $T_{\mu \nu}$ are denoted as $X := T_{\mu \nu} T^{\mu v}$ and $Y := T_{\mu \nu} \widetilde{T}^{\mu v}$, with the dual defined by $\widetilde{T}_{\mu \nu} := \frac{1}{2} \epsilon_{\mu v \alpha \beta} T^{\alpha \beta}$. 
Note that for vanishing coefficients for Lorentz violation, the dispersion relation Equation~(\ref{eq:smemdr}) reduces to the usual form $\left(p^2-m^2\right)^2=0$, which is effectively quadratic. The quadratic nature is retained when the only nonzero coefficients are $a_\mu, c_{\mu \nu}, e_\mu$, and $f_\mu$. However, the dispersion relation is generically quartic if $b_\mu, d_{\mu \nu}, g_{\lambda \mu \nu}$, or $H_{\mu \nu}$ are nonzero.

To determine the relativistic Lagrangian $L$ from an MDR $\mathcal{R}(p)=0$,
in addition to the workflow presented in Section~\ref{sec:cons-mdrF},
a method was developed in Ref.~\cite{kinematics-lv}, presented as five key equations:
\begin{equation}\label{eq:derivation}
\begin{split}
    &\mathcal{R}(p)=0,\\
    &y^j=-y^0\frac{\partial p_0}{\partial p_j}, j=1,2,3,\\
    &L=-p_\mu y^\mu.
\end{split}
\end{equation}
The first equation of Equation~(\ref{eq:derivation}) is the MDR we already have.
Recall that $y^\mu=\frac{d x^\mu}{d \tau}$ is the 4-velocity of the particle and $p_\mu :=-\frac{\partial L}{\partial y^\mu}$, 
the second equation of Equation~(\ref{eq:derivation}) means that the traditional relation 
\begin{equation}
    v_j=\frac{\partial E}{\partial p_j}
\end{equation}
holds in this situation, where $v_j=\frac{y^j}{y^0}=\frac{d x^j}{d x^0}=\frac{d x^j}{d t}$ is the speed of the particle.
The third equation of Equation~(\ref{eq:derivation}) is just from the homogeneity of $L$, as Equation~(\ref{eq:homo}).
The five equations can be manipulated to eliminate the 4-momentum components, leaving a single equation that can be viewed as a polynomial for $L$.

For the quadratic case of Equation~(\ref{eq:smemdr}), any form of MDR can be written as
\begin{equation}
    (p+\kappa)\Omega(p+\kappa)=\mu^2, 
\end{equation}
where $\mu>0$ is a mass-like scalar, $\kappa_\mu$ is a constant 4-vector shift of the momentum, and $\Omega^{\mu \nu}$ is a constant metric-like symmetric tensor. In the limit of vanishing background fields, $\Omega^{\mu v} \rightarrow \eta^{\mu \nu}$, $\kappa_\mu \rightarrow 0$, and $\mu \rightarrow m$. For perturbative background fields, $\Omega$ is invertible. The solution of Equation~(\ref{eq:derivation}) yields
\begin{equation}
    L=-\mu \sqrt{y\Omega^{-1}y}+\kappa\cdot y.
\end{equation}
For the special case of nonzero coefficients $a_\mu$, $c_{\mu\nu}$, $e_\mu$ and $,f_\mu$, $\Omega=\left(\delta+2 c+c^T c-e e-f f\right)$.
With additional condition $c_{\mu\nu}=0$, $(\Omega^{-1})_{\mu\nu}$ can be expressed in finite form and the corresponding Lagrangian is given by
\begin{adjustwidth}{-\extralength}{0cm}
\begin{equation}
\begin{aligned}
L= & -\mu\left\{y^2+\frac{1}{\Delta}\left[\left(1-f^2\right)(e \cdot y)^2+\left(1-e^2\right)(f \cdot y)^2+2(e \cdot f)(e \cdot y)(f \cdot y)\right]\right\}^{\frac{1}{2}}-a \cdot y \\
& +\frac{1}{\Delta}\left[\left(1-f^2\right)(m-e \cdot a)-(e \cdot f)(f \cdot a)\right] e \cdot y \\
& +\frac{1}{\Delta}\left[(e \cdot f)(m-e \cdot a)-\left(1-e^2\right)(f \cdot a)\right] f \cdot y,
\end{aligned}
\end{equation}
\end{adjustwidth}
where
\begin{equation}
\mu= \frac{1}{\sqrt{\Delta}}\left[\left(1-f^2\right)(m-e \cdot a)^2  -2(e \cdot f)(f \cdot a)(m-e \cdot a) +\left(1-e^2\right)(f \cdot a)^2\right]^{\frac{1}{2}}
\end{equation}
and
\begin{equation}
    \Delta =(1-e^2)(1-f^2)-(e\cdot f)^2
\end{equation}
is the determination of $\Omega$.
For the quartic cases when the coefficients $b_\mu$, $d_{\mu \nu}$, $H_{\mu \nu}$, or $g_{\lambda \mu\nu}$ are nonzero, things are more complicated.
When the first nonzero coefficients are $a_\mu$ and $b_\mu$, the MDR is
\begin{equation}
    \mathcal{R}(p)=\left[-(p-a)^2+b^2+m^2\right]^2-4\left[b\cdot(p-a)\right]^2+4b^2(p-a)^2=0,
\end{equation}
and the corresponding Lagrangian is
\begin{equation}\label{eq:Lagrangian_ab}
    L=-m \sqrt{y^2}-a \cdot y \mp \sqrt{(b \cdot y)^2-b^2 y^2}.
\end{equation}
When $b_\mu=0$, the Finsler geometry corresponding to the above Lagrangian is of Randers.
Another case with only $H_{\mu\nu}$ nonzero yields
\begin{equation}\label{eq:Lagrangian_H}
    L=-m \sqrt{y^2} \pm \sqrt{y H H y+2 X y^2}.
\end{equation}
The common feature of Equation~(\ref{eq:Lagrangian_ab}) and Equation~(\ref{eq:Lagrangian_H}) is that if $a_\mu=0$, the corresponding Finsler norms have the form
\begin{equation}
    F=\sqrt{y^2}\pm\sqrt{y^\mu s_{\mu\nu}y^\nu}.
\end{equation}
In this case, the Finsler geometry is called bipartite Finsler geometry. Detailed discussions of bipartite Finsler geometry can be found in Refs.~\cite{bipartite-sme, Silva2014, Colladay2017, Silva2019}.
For modified Dirac field theory at higher orders, it is impossible to write down the corresponding Lagrangians and Finsler norms explicitly. Instead, perturbative expansions for Lagrangians are adapted in Refs.~\cite{sme-higher-order1,sme-higher-order2}.

For the case of Lorentz-violating scalar fields, the effective quadratic Lagrange density of a complex scalar field $\phi(x^\mu)$ can be written in the form
\begin{equation}
\mathcal{L}\left(\phi, \phi^{\dagger}\right)=  \partial^\mu \phi^{\dagger} \partial_\mu \phi-m^2 \phi^{\dagger} \phi  -\frac{1}{2}\left(i \phi^{\dagger}\left(\hat{k}_a\right)^\mu \partial_\mu \phi+\text { h.c. }\right)+\partial_\mu \phi^{\dagger}\left(\hat{k}_c\right)^{\mu \nu} \partial_\nu \phi,
\end{equation}
where $(\hat{k}_a)^\mu$ and $(\hat{k}_c)^{\mu \nu}$ are operators constructed as series of even powers of the partial spacetime derivatives $\partial_\alpha$.
The MDR for the above Lagrange density is
\begin{equation}
    p^2-m^2-\left(\hat{k}_a\right)^\mu p_\mu+\left(\hat{k}_c\right)^{\mu \nu} p_\mu p_v=0,
\end{equation}
where 
the operators $\left(\hat{k}_a\right)^\mu$ and $\left(\hat{k}_c\right)^{\mu \nu}$ can conveniently be expressed as expansions in even powers of the $n$-momentum $p_\mu$ of the form
\begin{equation}
\begin{aligned}
&\left(\hat{k}_a\right)^\mu =\sum_{d \geq n-1}\left(k_a^{(d)}\right)^{\mu \alpha_1 \alpha_2 \ldots \alpha_{d-n+1}} p_{\alpha_1} p_{\alpha_2} \ldots p_{\alpha_{d-n+1}}, \\
&\left(\hat{k}_c\right)^{\mu \nu} =\sum_{d \geq n}\left(k_c^{(d)}\right)^{\mu \nu \alpha_1 \alpha_2 \ldots \alpha_{d-n}} p_{\alpha_1} p_{\alpha_2} \ldots p_{\alpha_{d-n}}.
\end{aligned}
\end{equation}
The explicit solution for the above MDR from Equation~(\ref{eq:derivation}) is also impossible.
The implicit solution~\cite{finsler-scalar} can be written in the form
\begin{equation}\label{eq:Lagrangian_scalar}
\begin{aligned}
L= & - \big{[}m^2 \bar{y}^2  \\
& +\frac{1}{16}\left(\sum_{d\geq n+1}(d-n+2) y_{\alpha_1} p_{\alpha_2} \ldots p_{\alpha_{d-n+2}}\left(k^{(d)}\right)^{\alpha_1 \ldots \alpha_{d-n+2}}\right)^2 \\
& +\frac{1}{2} \sum_{d\geq n+1}(d-n) \bar{y}^2 p_{\alpha_1} \ldots p_{\alpha_{d-n+2}}\left(k^{(d)}\right)^{\alpha_1 \ldots \alpha_{d-n+2}} \big{]}^{\frac{1}{2}} \\
& +\frac{1}{4} \sum_{d\geq n+1}(d-n+2) y_{\alpha_1} p_{\alpha_2} \ldots p_{\alpha_{d-n+2}}\left(k^{(d)}\right)^{\alpha_1 \ldots \alpha_{d-n+2}},
\end{aligned}
\end{equation}
with implicit relations
\begin{equation}
\begin{aligned}
y^\mu= & -L p_\nu\left[\eta^{\mu \nu}+\frac{1}{2} \sum_{d\geq n+1}(d-n+2) p_{\alpha_1} \ldots p_{\alpha_{d-n}}\left(k^{(d)}\right)^{\alpha_1 \ldots \alpha_{d-n} \mu \nu}\right] \\
& \times\left[m^2+\frac{1}{2} \sum_{d\geq n+1}(d-n) p_{\alpha_1} \ldots p_{\alpha_{d-n+2}}\left(k^{(d)}\right)^{\alpha_1 \ldots \alpha_{d-n+2}}\right]^{-1}
\end{aligned}
\end{equation}
and $\bar{y}^2=\sqrt{\eta_{\mu\nu} y^\mu y^\nu}$.
The expansion solution of Equation~(\ref{eq:Lagrangian_scalar}) can be found in Ref.~\cite{finsler-scalar}.

\subsection{Very Special Relativity and Finsler Geometry}\label{sec:VSR}

Very special relativity has roots in the early research of Bogoslovsky.
By the year 1977, to explain the discrepancy between the Greisen–Zatsepin–Kuzmin cutoff (GZK cutoff) and the experimental data, 
Bogoslovsky suggested the violation of usual relativistic relations at Lorentz  {factors} 
 $\gamma \geq 5\times 10^{10}$~\cite{Bogoslovsky1977}.
In calculating the point of the spectrum cutoff, the Lorentz transformation from the Earth-centered frame of reference to that of the proton at rest is essentially used, but the relative velocity of these coordinate systems is close to the velocity of light and it is far from being evident that the formula of the special relativity theory remains valid up to velocities infinitely close to the velocity of light at that time.
He suggested the new transformation law at velocities very close to the velocity of light is essentially different from the old one.
Following Ref.~\cite{Bogoslovsky1977}, to generalize the Lorentz transformation written in 
\begin{equation}
    \left\{\begin{array}{l}
x_0^{\prime}=x_0 \cosh \alpha-x \sinh \alpha, \\
x^{\prime}=-x_0 \sinh \alpha+x \cosh \alpha,
\end{array} \right.
\end{equation}
where $\tanh \alpha=V$,
he assumed that the wave equation $\left(\partial^2 / \partial x_0^2-\partial^2 / \partial x^2\right) \varphi=0$ is invariant under the new transformation law.
Thus, he wrote the generalized transformations
\begin{equation}\label{eq:transVSR}
    \left\{\begin{array}{l}
x_0^{\prime}=\exp [-r \alpha]\left(x_0 \cosh \alpha-x \sinh \alpha\right), \\
x^{\prime}=\exp [-r \alpha]\left(-x_0 \sinh \alpha+x \cosh \alpha\right),
\end{array}\right.
\end{equation}
where $r$ is a dimensionless parameter of scale transformations.
Equation~(\ref{eq:transVSR}) can be written in the form
\begin{equation}
\left\{\begin{array}{l}
x_0^{\prime}=\left(\frac{1-V }{1+V}\right)^{r / 2} \frac{x_0-V  x}{\sqrt{1-V^2}}, \\
x^{\prime}=\left(\frac{1-V }{1+V }\right)^{r / 2} \frac{x- /  x_0}{\sqrt{1-V^2 }}.
\end{array}\right.
\end{equation}
By writing Equation~(\ref{eq:transVSR}) in infinitesimal form, 
the invariant function
\begin{equation}
    f(x_0,x)=\left(\frac{x_0-x}{x_0+x}\right)^r\left(x_0^2-x^2\right)
\end{equation}
was found; thus, the new differential invariant is
\begin{equation}
    \mathrm{d} s=\left(\frac{\mathrm{d} x_0-\mathrm{d} x}{\mathrm{~d} x_0+\mathrm{d} x}\right)^{r/ 2} \sqrt{\mathrm{d} x_0^2-\mathrm{d} x^2},
\end{equation}
and in four-dimensional spacetime,  {the generalized form is} 
\begin{equation}\label{eq:finsler-Bogoslovsky}
    \mathrm{d} s=\left[\frac{\left(\mathrm{d} x_0-\vec{v}\cdot \mathrm{d} \vec{x}\right)^2}{\mathrm{~d} x_0^2-\mathrm{d} \vec{x}^2}\right]^{r / 2} \sqrt{\mathrm{d} x_0^2-\mathrm{d} \vec{x}^2},
\end{equation}
where $|\vec{v}|=1$.
\textls[-15]{In fact, Equation~(\ref{eq:finsler-Bogoslovsky}) is just a Finsler norm.
The corresponding relativistic kinematics and Finsler geometries were studied by Bogoslovsky and Goenner later~\cite{Bogoslovsky1998, Bogoslovsky1999-1, Bogoslovsky1999-2}.}
The group theories of the above spacetime were also studied in Ref.~\cite{Bogoslovsky1977}.

On the other hand, in 2006, Cohen and Glashow pursued a different approach to the possible failure of Lorentz symmetry, using group theory.
They first introduced the very special relativity~\cite{vsr}, 
which means descriptions of nature whose spacetime symmetries are certain proper subgroups of the Poincaré group.
These subgroups contain spacetime translations together with at least a two-parameter subgroup of the Lorentz group isomorphic to that generated by
$T_1:=K_x+J_y$ and $T_2:=K_y-J_x$,
where $\mathbf{J}$ and $\mathbf{K}$ are the generators of rotations and boosts, respectively.
These commuting generators form a group, $T(2)$, which is isomorphic to the group of translations in the plane.
$T_1, T_2$ with the addition of $J_z$ yields a group isomorphic to the three-parameter group of Euclidean motions, $E(3)$.
$T_1, T_2$ with the addition of $K_z$ yields one isomorphic to the three-parameter group of orientation-preserving similarity transformations, or homotheties, $HOM(2)$.
The addition of both $J_z$ and $K_z$ yields one isomorphic to the four-parameter similitude group, $SIM(2)$.
The corresponding Lie algebras and their Lie brackets are listed in Table~\ref{tab:vsr}~\cite{Ahluwalia2010}.
Usually, the special relativity with symmetry $SIM(2)$ is referred to as VSR.

\begin{table}[H] 
\caption{ {The} 
 four VSR algebras.\label{tab:vsr}}
\newcolumntype{x}[1]{>{\centering\arraybackslash}p{#1}}
\begin{tabularx}{\textwidth}{x{3cm} x{3cm} x{5cm}}
\toprule
\textbf{Designation}	& \textbf{Generators}	& \textbf{Algebra}\\
\midrule
$\mathfrak{t}(2)$     & $T_1, T_2$            & ${\left[T_1, T_2\right]=0}$ \\
$\mathfrak{e}(2)$     & $T_1, T_2, J_z$       & ${\left[T_1, T_2\right]=0,\left[T_1, J_z\right]=-i T_2,\left[T_2, J_z\right]=i T_1}$ \\
$\mathfrak{hom}(2)$   & $T_1, T_2, K_z$       & ${\left[T_1, T_2\right]=0,\left[T_1, K_z\right]=i T_1,\left[T_2, K_z\right]=i T_2}$ \\
$\mathfrak{sim}(2)$   & $T_1, T_2, J_z, K_z$  & ${\left[T_1, T_2\right]=0,\left[T_1, K_z\right]=i T_1,\left[T_2, K_z\right]=i T_2},$ \\
                      &                       & ${\left[T_1, J_z\right]=-i T_2,\left[T_2, J_z\right]=i T_1,\left[J_z, K_z\right]=0}$ \\
\bottomrule
\end{tabularx}
\end{table}

The semidirect product between the translations 
and $SIM(2)$ gives an eight-dimensional subgroup of the Poincaré group called $ISIM(2)$,
which was first introduced but not named in 1970 by Kogut and Soper~\cite{Kogut1970}.
In 2007, Gibbons  {et~al.}~\cite{vsr-finsler} tried to deform $ISIM(2)$ similar to deforming the Poincaré group into de Sitter (anti-de Sitter) group~\cite{Monique1967}, with noncommutative relation $\left[P_\mu, P_\nu\right]=\frac{1}{3}\Lambda M_{\mu,\nu}$, where $\Lambda$ is the cosmological constant.
They obtained a one-parameter family of deformations, denoted by $DISIM_b(2)$. 
Different from de Sitter (anti-de Sitter) group, the translations of $DISIM_b(2)$ remain commutative. The nontrivial Lie brackets for the algebra $\mathfrak{disim}_b(2)$ are given by
\begin{equation}
\begin{aligned}
& {\left[N, P_{ \pm}\right]=-(b \pm 1) P_{ \pm}, \quad\left[N, P_i\right]=-b P_i \text {, }} \\
& {\left[N, M_{+i}\right]=-M_{+i}, \quad\left[J, P_i\right]=\epsilon_{i j} P_j,} \\
& {\left[J, M_{+i}\right]=\epsilon_{i j} M_{+i}, \quad\left[M_{+i}, P_{-}\right]=P_j \text {, }} \\
& {\left[M_{+i}, P_j\right]=-\delta_{i j} P_{+} ,} \\
\end{aligned}
\end{equation}
where $N:=M_{+-}$ and $J:=M_{12}$, with the Minkowski metric to be $ds^2=\eta_{\mu\nu}x^\mu x^\nu=2dx^+dx^-+dx^idx^i$, and i and j ranging over the values 1 and 2.
The deformed generator $N$ acts as
\begin{equation}
    x^i \rightarrow \lambda^{-b} x^i, \quad x^{-} \rightarrow \lambda^{1-b} x^{-}, \quad x^{+} \rightarrow \lambda^{-1-b} x^{+};
\end{equation}
thus, the invariant line element is a Finsler line element
\begin{equation}\label{eq:finsler-vsr}
\begin{aligned}
d s & =\left(2 d x^{+} d x^{-}+d x^i d x^i\right)^{(1-b) / 2}\left(d x^{-}\right)^b \\
& =\left(\eta_{\mu \nu} d x^\mu d x^\nu\right)^{(1-b) / 2}\left(n_\rho d x^\rho\right)^b.
\end{aligned}
\end{equation}
Equation~(\ref{eq:finsler-vsr}) is almost the same as Equation~(\ref{eq:finsler-Bogoslovsky}), found by Bogoslovsky in 1977 in different ways.
Thus, the general very special relativity is Finsler geometry, as per the title of Ref.~\cite{vsr-finsler}.
The Finsler norm of the $DISIM_b(2)$ VSR spacetime from Equation~(\ref{eq:finsler-vsr}) is
\begin{equation}
    F=(-\eta_{\mu\nu} y^\mu y^\nu )^{(1-b)/2}(-n_\rho y^\rho)^b,
\end{equation}
and the Lagrangian is
\begin{equation}
    L=-m\left(-\eta_{\mu \nu} \dot{x}^\mu \dot{x}^\nu\right)^{(1-b) / 2}\left(-n_\rho \dot{x}^\rho\right)^b.
\end{equation}
The dispersion relation of the above Lagrangian is
\begin{equation}
    \eta^{\mu \nu} p_\mu p_\nu=-m^2\left(1-b^2\right)\left(-\frac{n^\nu p_\nu}{m(1-b)}\right)^{2 b /(1+b)},
\end{equation}
and the corresponding Klein--Gordon equation is
\begin{equation}
    -\square \phi+m^2\left(1-b^2\right)\left(\frac{\mathrm{i} n^\mu \partial_\mu}{m(1-b)}\right)^{2 b /(1+b)} \phi=0 .
\end{equation}
The Finsler structure of the $DISIM_b(2)$ VSR is quite different from the Finsler structures in Section~\ref{sec:mdrF}, as there is no LV energy scale in VSR.

For the case of $SIM(2)$ VSR, one interesting result is that it may provide an origin of Lorentz violent neutrino mass~\cite{vsr-neutrino}.
Depending on Ref.~\cite{vsr-neutrino}, a $SIM(2)$-invariant modification of the electrodynamic theory is established~\cite{Cheon2009}.
The $SIM(2)$-covariant Dirac equation can be written in the form
\begin{equation}\label{eq:sim2Dirac}
    \left[i \gamma^\mu\left(\partial_\mu+\frac{\lambda}{2} N_\mu\right)-m\right] \Psi(x)=0,
\end{equation}
where $N_\mu:=\frac{n_\mu}{n\cdot \partial}$ with a chosen preferred null direction $n^\mu=(1,0,0,1)$ and attributes $N\cdot N=0, N\cdot \partial=1$.
Squaring Equation~(\ref{eq:sim2Dirac}), one obtains
\begin{equation}
    \left[\partial^\mu \partial_\mu+M^2\right] \Psi(x)=0,
\end{equation}
where $M^2=m^2+\lambda$. 
Thus, the physical mass $M$ does not vanish if $\lambda \ne 0$, even with $m=0$.
The field theory of symmetry $SIM(2)$ and the Very Special Relativity Standard Model were developed by Alfaro  {et~al.}~\cite{Alfaro2013, Alfaro2014, Alfaro2015, Alfaro2017, Alfaro2018, Alfaro2019, Alfaro2019-2, Alfaro2019-3, Alfaro2021, Alfaro2022, Haghgouyan2022, Bufalo2022, Cheng-Yang2015, Ilderton2016} later.
Interestingly, in the VSR field theory, a gauge-invariant photon mass~\cite{Alfaro2019-2} and graviton mass~\cite{Alfaro2022} are allowed.

For the case of $DISIM_b(2)$ VSR, the field theory is more complicated. 
For photons, a mass term is suggested, given as~\cite{vsr-finsler}
\begin{equation}
    L=-\frac{1}{4} F_{\mu \nu} F^{\mu \nu}-\frac{1}{2} m^2\left(\frac{\left(n^\mu A_\mu\right)^2}{A_\nu A_\nu}\right)^b A_\rho A^\rho,
\end{equation}
and the mass term of the Dirac Lagrangian is suggested by Bogoslovsky~\cite{Bogoslovsky2004} as
\begin{equation}
    m\left[\left(\frac{\mathrm{i}n_\mu \bar{\psi} \gamma^\mu \psi}{\bar{\psi} \psi}\right)^2\right]^{b / 2} \bar{\psi} \psi,
\end{equation}
where $n_\mu$ is the same as in Equation~(\ref{eq:finsler-vsr}).

The VSR in curved spacetime also arouses people's interest.
For the case of $SIM(2)$ VSR, the VSR to the physical situation in which a cosmological constant is present is considered by Alvarez and Vidal~\cite{Alvarez2008}, 
and the generalization of $SIM(2)$ VSR to curved spacetimes is  {considered by} 
 Mück~\cite{Muck2008}.
For the case of $DISIM_b(2)$ VSR, since it is connected with Finsler spacetime in the form
\begin{equation}
    F(x, y)=\left(\eta_{\mu\nu} y^\mu y^\nu\right)^{(1-b) / 2}\left(n_\rho y^\rho\right)^b,
\end{equation}
geometric methods can be applied.
The Finsler cosmology of the $DISIM_b(2)$ VSR was considered by Kouretsis  {et~al.}~\cite{Kouretsis2009}.
In their consideration, they replace the Minkowskian metric $\eta_{\mu\nu}$ in Equation~(\ref{eq:finsler-vsr}) with the FRW one 
\begin{equation}
    a_{\mu \nu}=\operatorname{diag}\left(1,-\frac{a^2(t)}{1-k r^2},-a^2(t) r^2,-a^2(t) r^2 \sin ^2 \theta\right),
\end{equation}
and they consider the case of $n^\mu=(n(t),0,0,0)$ with $|n^\mu|\ll1$ and Taylor expansion $n(t)\sim At+B$.
In this case, the  Friedman equation is reconsidered and an extra term $\Omega_X=-2 \frac{A}{B H} b$ is present and might give a significant
contribution to the acceleration compared to the dark energy parameter $\Omega_\Lambda$.
In addition, an additional redshift effect due to the Lorentz violations inherited by the parameter $b$ and the spurion vector field $n_\mu$ is present, and the energy of the massive nonrelativistic particle will change with time as 
\begin{equation}
    E=E_0 \exp \left(-\frac{2 A b}{B} t\right).
\end{equation}
The very special relativity in curved spacetime is of Berwald--Finsler spacetime was also studied by Fuster  {et~al.}~\cite{Fuster2018}.

\section{Summary and Discussion}\label{sec:discussion}

As a generalization of Riemann geometry, Finsler geometry provides more possibilities for new physics.
Different from Riemann geometry, the square of the length element $ds^2$ in Finsler geometry is not necessarily a quadratic form of $dx^\mu$,
and thus Finsler norms can be of infinite choices of forms. 
Unfortunately or luckily, there are no criteria to constrain what kind of Finsler geometries should be applied in physics,
so research on Finsler geometries applied in physics springs up in different areas, as per the examples shown in Section~\ref{sec:introduction}.

The application of Finsler geometry in LV studies is a success.
On the one hand, theories on LV, if accompanied by modified dispersion relations, can be connected with Finsler geometries via the methodology in Section~\ref{sec:cons-mdrF}.
On the other hand, theories without LV energy scales may also be connected to Finsler geometries via spacetime symmetries, such as VSR in Section~\ref{sec:VSR}.
Finsler geometry provides descriptions of spacetime for LV theories and has advantages in kinematics research.
In Section~\ref{sec:physicsF}, we reviewed the effects of time dilation and arrival time delay formula in the Finsler geometry description of LV.
With the assumption of the principle of relativity, we also presented the transformations between inertial frames and the modified composition laws. These modified composition laws together with the MDRs may bring different views on threshold anomalies.
Finsler geometry also shows the possibility of viewing other LV theories from different perspectives.
One proposal, proposed by Magueijo and Smolin ~\cite{DSR2-2}, to address the difficulty of applying DSR to macroscopic objects,
is replacing $E_{\rm DSR}$ with $N E_{\rm DSR}$ when a  macroscopic system has $N$ particles.
This proposal is just an assumption in Ref.~\cite{DSR2-2}, but in Finsler geometry, it is a result of the analysis of the Finsler structure, as per the discussion in Section~\ref{sec:DSR}.

Maybe we could progress further by assuming that the Finsler structure from LV is an attribute of the universe, and that means that all particles and objects in the universe have 
the same Finsler structure. Let us assume that the MDRs of these objects are
\begin{equation}
    E^2=m^2+p^2+\frac{s}{E_{\rm LV}^n}E^{n+2} G\left(\frac{\Vec{p}}{E}\right),
\end{equation}
where $s=\pm 1$, $E_{\rm LV}$ is the LV scale, and $G(x)$ is an abstract function.
The corresponding Finsler norms are
\begin{equation}
    F(y)=\sqrt{y^2}+\frac{s}{2}\left(\frac{m}{E_{\rm LV}}\right)^n\frac{(y^0)^{n+2}}{\left(\sqrt{y^2}\right)^{n+1}}G\left(-\frac{\Vec{y}}{y^0}\right),
\end{equation}
where $y^2=(y^0)^2-\delta_{ij}y^iy^j$ and $i,j$ are spatial indexes. 
The assumption on the Finsler structures suggests that $m/E_{\rm LV}$ is a constant for all objects, or
\begin{equation}
    E_{\rm LV, p}=C m_{\rm p},  
\end{equation}
where different particles are identified by the index $\rm p$ and $C$ is a dimensionless constant of the universe.  
This means that for massive particles, the LV scales for different particles are proportional to their masses, 
and the heavier the particle is, the higher the LV scale is and the less obvious the LV effects are.
Of course, this inference is too radical, but it conforms to the phenomenological cognition of LV.
Or, we can assume the weak version of the inference, that the LV scales of particles and the masses of particles are positively correlated for massive particles. 
Of course, more tests are needed to examine the assertion.

Although Finsler geometry performs successfully in the studies of LV, it is not advisable to regard Finsler geometry as the final theory of LV.
It works well when dealing with spacetime and gravity, but it lacks the description of other interactions.
Another challenge is that we lack a methodology to construct the Finsler norms that describes LV directly based on some basic assumptions.
Worse still, we cannot construct Finsler norms from MDRs for massless particles.
These problems mean that, at present, Finsler geometry can be considered an effective theory when dealing with LV, and more research is needed in applying Finsler geometry in LV studies.

\vspace{6pt} 



\authorcontributions{Both J.Z. and B.-Q.M. 
contributed extensively to 
this project. All authors have
read and agreed to the published version of the manuscript.}

\funding{This work was supported by the National Natural Science Foundation of China,
grant number 12075003.}

\conflictsofinterest{The authors declare no conflicts of interest.}

\institutionalreview{Not applicable.}

\informedconsent{Not applicable.}

\dataavailability{No new data were created or analyzed in this study. Data sharing is not applicable to this article.}




\abbreviations{Abbreviations}{
The following abbreviations are used in this manuscript:\\

\noindent 
\begin{tabular}{@{}ll}
DSR & Doubly special relativity\\
FRW & Friedmann–Lemaître–Robertson–Walker\\
GZK &  Greisen–Zatsepin–Kuzmin\\
LV  & Lorentz violation\\
MDR & Modified dispersion relation\\
QG  & Quantum gravity\\
SME & Standard-Model Extension\\
VSR & Very special relativity

\end{tabular}
}




\begin{adjustwidth}{-\extralength}{0cm}

\reftitle{References}

\PublishersNote{}
\end{adjustwidth}

\begin{thebibliography}{999}



\bibitem[Amelino(2005)]{qg}
Amelino-Camelia, G.
Introduction to Quantum-Gravity Phenomenology.
{\em Lect. Notes Phys.} {\bf 2005}, {\em 669}, 59.

\bibitem[Danielsson(2001)]{string}
Danielsson, U.
Introduction to string theory.
{\em Rep. Prog. Phys.} {\bf 2001}, {\em 64}, 51.

\bibitem[Li(2021)]{Chengyi2021}
Li, C.; Ma, B.-Q.
Light speed variation in a string theory model for space-time foam.
{\em Phys. Lett. B} {\bf 2021}, {\em 819}, 136443.

\bibitem[Li(2022)]{Chengyi2022}
Li, C.; Ma, B.-Q.
Lorentz- and CPT-violating neutrinos from string/D-brane model.
{\em Phys. Lett. B} {\bf 2022}, {\em 835}, 137543.

\bibitem[Li(2023)]{Chengyi2023}
Li, C.; Ma, B.-Q.
Lorentz and CPT breaking in gamma-ray burst neutrinos from string theory.
{\em  {J. High Energy Phys.}} {\bf 2023}, {\em  {03}}, 230.   





\bibitem[Rovelli(2008)]{Rovelli2008}
Rovelli, C.
Loop quantum gravity.
{\em Living Rev. Relativ.} {\bf 2008}, {\em 11}, 5.

\bibitem[Ashtekar(2021)]{Ashtekar2021}
Ashtekar, A.; Bianchi, E.
A short review of loop quantum gravity.
{\em Rep. Prog. Phys.} {\bf 2021}, {\em 84}, 042001.

\bibitem[Li(2023)]{lihao2023}
Li, H.; Ma, B.-Q.
Speed variations of cosmic photons and neutrinos from loop quantum gravity.
{\em Phys. Lett. B} {\bf 2013}, {\em 836}, 137613.



\bibitem[Amelino(2001)]{DSR1-1}
Amelino-Camelia, G.
Testable scenario for relativity with minimum length. 
{\em Phys. Lett. B} {\bf 2001}, {\em 510}, 255–263.

\bibitem[Amelino(2002)]{DSR1-2}
Amelino-Camelia, G.
Relativity in space-times with short-distance structure governed by an observer-independent (Planckian) length scale. 
{\em Int. J. Mod. Phys. D} {\bf 2002}, {\em 11}, 35.

\bibitem[Smolin(2002)]{DSR2-1}
Magueijo, J.; Smolin, L.
Lorentz invariance with an invariant energy scale. 
{\em Phys. Rev. Lett. } {\bf 2002}, {\em 88}, 190403.

\bibitem[Smolin(2003)]{DSR2-2}
Magueijo, J.; Smolin, L.
Generalized Lorentz invariance with an invariant energy scale. 
{\em Phys. Rev. D} {\bf 2003}, {\em 67}, 044017.

\bibitem[Glashow(2006)]{vsr}
Cohen, A.G.; Glashow, S.L. 
Very Special Relativity. 
{\em Phys. Rev. Lett.} {\bf 2006}, {\em 97}, 021601.

\bibitem[Colladay(1997)]{cpt-sm}
Colladay, D.; Kostelecký, V.A. 
$CPT$ violation and the standard model. 
{\em Phys. Rev. D} {\bf 1997}, {\em 55}, 6760.

\bibitem[Colladay(1998)]{lv-sm}
Colladay, D.; Kostelecký, V.A. 
Lorentz-violating extension of the standard model. 
{\em Phys. Rev. D} {\bf 1998}, {\em 58}, 116002.

\bibitem[ Kostelecký(2004)]{glv-sm}
Kostelecký, V.A. 
Gravity, Lorentz violation, and the standard model. 
{\em Phys. Rev. D} {\bf 2004}, {\em 69}, 105009.

\bibitem[He(2022)]{HeMa}
He, P.; Ma, B.-Q.
Lorentz Symmetry Violation of Cosmic Photons.
{\em Universe} {\bf 2022}, {\em 8}, 323.

\bibitem[ChangZ(2008)]{ChangZ2008}
Chang, Z.; Li, X.
Lorentz invariance violation and symmetry in Randers-Finsler spaces.
{\em Phys. Lett. B} {\bf 2008}, {\em 663}, 103--106.

\bibitem[ChangZ(2008)]{ChangZ2009}
Chang, Z.; Li, X.
Ultra-high energy cosmic rays threshold in Randers-Finsler space.
{\em Chin. Phys. C} {\bf 2009}, {\em 33},  626--628.


\bibitem[LiX(2010)]{LiX2010}
Li, X.; Chang, Z.
Towards a gravitation theory in Berwald–Finsler space.
{\em Chin. Phys. C} {\bf 2010}, {\em 34}, 28.

\bibitem[Pfeifer(2012)]{Pfeifer2012}
Pfeifer, C.; Wohlfarth, M.N.R.
Finsler geometric extension of Einstein gravity
{\em Phys. Rev. D} {\bf 2012}, {\em 85}, 064009.

\bibitem[LiX(2013)]{LiX2013}
Li, X.; Chang, Z.
Spacetime structure of MOND with Tully-Fisher relation and Lorentz invariance violation.
{\em Chin. Phys. C} {\bf 2013}, {\em 37}, 123103.

\bibitem[ChangZ(2008)]{ChangZ2008-2}
Chang, Z.; Li, X.
Modified Newton's gravity in Finsler space as a possible alternative to dark matter hypothesis.
{\em Phys. Lett. B} {\bf 2008}, {\em 668}, 453--456.

\bibitem[Saridakis(2013)]{Saridakis2013}
  Basilakos, S.; Kouretsis, A.P.; Saridakis, E.N.; Stavrinos, P.C.
Resembling dark energy and modified gravity with Finsler-Randers cosmology.
{\em Phys. Rev. D} {\bf 2013}, {\em 88}, 123510.

\bibitem[ChangZ(2008)]{ChangZ2009-2}
Chang, Z.; Li, X.
Modified Friedmann model in Randers–Finsler space of approximate Berwald type as a possible alternative to dark energy hypothesis.
{\em Phys. Lett. B} {\bf 2009}, {\em 676}, 173--176.

\bibitem[Saridakis(2019)]{Saridakis2019-1}
Minas, G.; Saridakis, E.N.; Stavrinos, P.C.; Triantafyllopoulos, A. 
Bounce Cosmology in Generalized Modified Gravities.
{\em Universe} {\bf 2019}, {\em 5}, 74.


\bibitem[ChangZ(2014)]{ChangZ2014}
 Chang, Z.; Li, X.; Lin, H.-N.; Wang, S.
Constraining anisotropy of the universe from different groups of type-Ia supernovae.
{\em Eur. Phys. J. C} {\bf 2014}, {\em 74}, 2821.

\bibitem[ChangZ(2014)]{ChangZ2014-2}
 Chang, Z.; Li, X.; Lin, H.-N.; Wang, S.
Constraining the anisotropy of the universe from supernovae and gamma-ray bursts.
{\em Mod. Phys. Lett. A} {\bf 2014}, {\em 29}, 1450067.




\bibitem[Saridakis(2019)]{Saridakis2019-2}
  Ikeda, S.; Saridakis, E.N.; Stavrinos, P.C.; Triantafyllopoulos, A.
Cosmology of Lorentz fiber-bundle induced scalar-tensor theories.
{\em Phys. Rev. D} {\bf 2019}, {\em 100}, 124035.

\bibitem[Saridakis(2021)]{Saridakis2021}
  Konitopoulos, S.; Saridakis, E.N.; Stavrinos, P.C.; Triantafyllopoulos, A.
Dark gravitational sectors on a generalized scalar-tensor vector bundle model and cosmological applications.
{\em Phys. Rev. D} {\bf 2021}, {\em 104}, 064018.

\bibitem[LiX(2013)]{LiX2013-2}
Li, X.; Chang, Z.
Gravitational Wave in Lorentz Violating Gravity.
{\em Commun. Theor. Phys.} {\bf 2013}, {\em 60}, 535--540.

\bibitem[Antonelli(2018)]{Antonelli2018}
Antonelli, V.; Miramonti, L.; Torri, M.D.C.
Neutrino oscillations and Lorentz invariance violation in a Finslerian geometrical model.
{\em Eur. Phys. J. C} {\bf 2018}, {\em 78}, 667.



\bibitem[Bao(2000)]{textbook}
Bao, D.; Chern, S.S.; Shen, Z.
\textit{An Introduction to Riemann–Finsler Geometry (Graduate Texts in Mathematics, 200)};
 Springer: New York,  {NY, USA,} 
 2000.

\bibitem[Kimberly(2004)]{Kimberly2004}
Kimberly, D.; Magueijo, J.; Medeiros, J.
Nonlinear relativity in position space. 
{\em Phys. Rev. D} {\bf 2004}, {\em 70}, 084007.

\bibitem[Magueijo(2004)]{rainbow}
Magueijo, J.; Smolin, L.
Gravity's rainbow. 
{\em Class. Quant. Grav.} {\bf 2004}, {\em 21}, 1725.

\bibitem[Magueijo(2004)]{Ellis2004}
Ellis, J.R.; Mavromatos, N.E.; Nanopoulos, D.V.
Derivation of a vacuum refractive index in a stringy space–time foam model.
{\em Phys. Lett. B} {\bf 2008}, {\em 31}, 412--417.

\bibitem[Girelli(2007)]{Girelli2007}
Girelli, F.; Liberati, S.; Percacci, R.; Rahmede, C.
Modified dispersion relations from the renormalization group of gravity. 
{\em Class. Quant. Grav.} {\bf 2007}, {\em 24}, 3995.

\bibitem[Aloisio(2006)]{Aloisio2006}
  Aloisio, R.; Galante, A.; Grillo, A.; Liberati, S.; Luzio, E.; Méndez, F.
Deformed special relativity as an effective theory of measurements on quantum gravitational backgrounds. 
{\em Phys. Rev. D} {\bf 2006}, {\em 73}, 045020.



\bibitem[Girelli(2007)]{Girelli2007-mdr}
Girelli, F.; Liberati, S.; Sindoni, L.
Planck-scale modified dispersion relations and Finsler geometry. 
{\em Phys. Rev. D} {\bf 2007}, {\em 75}, 064015.

\bibitem[Amelino(2014)]{dsrFinsler}
Amelino-Camelia, G.; Barcaroli, L.; Gubitosi, G.;
Realization of doubly special relativistic symmetries in Finsler geometries. 
{\em Phys. Rev. D} {\bf 2014}, {\em 90}, 125030.

\bibitem[Lobo(2017)]{Lobo2017-1}
Lobo, I.P.; Loret, N.; Nettel, F.
Investigation of Finsler geometry as a generalization to curved spacetime of Planck-scale-deformed relativity in the de Sitter case. 
{\em Phys. Rev. D} {\bf 2017}, {\em 95}, 046015.

\bibitem[Lobo(2017)]{Lobo2017-2}
Lobo, I.P.; Loret, N.; Nettel, F.
Rainbows without unicorns: Metric structures in theories with modified dispersion relations. 
{\em Eur. Phys. J. C} {\bf 2017}, {\em 77}, 451.

\bibitem[Lobo(2021)]{Lobo2021}
Lobo, I.P.; Pfeifer, C.
Reaching the Planck scale with muon lifetime measurements. 
{\em Phys. Rev. D} {\bf 2021}, {\em 103}, 106025.

\bibitem[JieZ(2022)]{JieZ2022}
Zhu, J.; Ma, B.-Q.
Lorentz-violation-induced arrival time delay of astroparticles in Finsler spacetime. 
{\em Phys. Rev. D} {\bf 2022}, {\em 105}, 124069.

\bibitem[Lobo(2022)]{Lobo2022}
Lobo, I.P.; Pfeifer, C.; Morais, P.H.; Batista, R.A.; Bezerra, V.B.
Finite Planck-scale-modified relativistic framework in Finsler geometry.  {In Proceedings of the Corfu Summer Institute 2021 ``School and Workshops on Elementary Particle Physics and Gravity'', Corfu, Greece, 29 August--9 October 2021; p. 334.} 




\bibitem[Ellis(2003)]{Ellis2003}
Ellis, J.R.; Mavromatos, N.E.; Nanopoulos, D.V.; Sakharov, A.S. 
Quantum-gravity analysis of gamma-ray bursts using wavelets.
{\em Astron. Astrophys.} {\bf 2003}, {\em 402}, 409–424.

\bibitem[Amelino(2003)]{Amelino2003}
Amelino-Camelia, G.
Proposal of a second generation of quantum-gravity-motivated Lorentz-symmetry tests: Sensitivity to effects suppressed quadratically by the Planck scale
{\em Int. J. Mod. Phys. D} {\bf 2003}, {\em 12}, 1633–1640. 


\bibitem[Ellis(2006)]{Ellis2006}
Ellis, J.R.; Mavromatos, N.E.; Nanopoulos, D.V; Sakharov, A.S.; Sarkisyan, E.K.G.
Robust limits on Lorentz violation from gamma-ray bursts.
{\em Astropart. Phys.} {\bf 2006}, {\em 25},  402–411.
Erratum in {\em Astropart. Phys.} {\bf 2008}, {\em 29},  158–159.

\bibitem[Jacob(2007)]{Jacob2007}
Jacob, U.; Piran, T.
Neutrinos from gamma-ray bursts as a tool to explore quantum-gravity-induced Lorentz violation.
{\em Nat. Phys.} {\bf 2007}, {\em 3}, 87–90. 

\bibitem[Biesiada(2009)]{Biesiada2009}
Biesiada, M.; Piórkowska, A. 
Lorentz invariance violation-induced time delays in GRBs in different cosmological models. 
{\em Class. Quantum Gravity} {\bf 2009}, {\em 26}, 125007.


\bibitem[Shao(2010)]{Shao2010-1}
Shao, L.; Xiao, Z.; Ma, B.-Q.
 {Lorentz violation from cosmological objects with very high energy photon emissions.}
{\em Astropart. Phys.} {\bf 2010}, {\em 33}, 312–315.   


\bibitem[Shao(2010)]{Shao2010-2}
Shao, L.; Ma, B.-Q.
Lorentz violation effects on astrophysical propagation of very high energy photons.
{\em Mod. Phys. Lett. A} {\bf 2010}, {\em 25}, 3251–3266.

\bibitem[fermi(2013)]{fermi2013}
Vasileiou, V.; Jacholkowska, A.; Piron, F.; Bolmont, J.; Couturier, C.; Granot, J.; Stecker, F.W.; Cohen-Tanugi, J.; Longo, F.
Constraints on Lorentz invariance violation from Fermi-Large Area Telescope observations of gamma-ray bursts.
{\em Phys. Rev. D} {\bf 2013}, {\em 87}, 122001 .


\bibitem[Zhang(2015)]{Zhang2015}
Zhang, S.; Ma, B.-Q.
 {Lorentz violation from gamma-ray bursts.}
{\em Astropart. Phys.} {\bf 2015}, {\em 61}, 108–112.

\bibitem[Pan(2015)]{Pan2015}
Pan, Y.; Gong, Y.; Cao, S.; Gao, H.; Zhu, Z.-H.
Constraints on the Lorentz invariance violation with gamma-ray bursts via a Markov Chain Monte Carlo approach.
{\em Astrophys. J.} {\bf 2015}, {\em 808}, 78.


\bibitem[Xu(2016)]{Xu2016-1}
Xu, H.; Ma, B.-Q.
Light speed variation from gamma-ray bursts.
{\em Astropart. Phys.} {\bf 2016}, {\em 82},  72–76.

\bibitem[Xu(2016)]{Xu2016-2}
Xu, H.; Ma, B.-Q.
 Light speed variation from gamma ray burst GRB 160509A.
{\em Phys. Lett. B } {\bf 2016}, {\em 760}, 602–604.

 \bibitem[Zou(2018)]{Zou2018}
Zou, X.-B.; Deng, H.-K.;  Yin, Z.-Y.; Wei, H.
Model-independent constraints on Lorentz invariance violation via the cosmographic approach. 
{\em Phys. Lett. B} {\bf 2018}, {\em 776}, 284–294.

\bibitem[Liu(2018)]{Liu2018}
Liu, Y.; Ma, B.-Q.
Light speed variation from gamma ray bursts: Criteria for low energy photons. 
{\em Eur. Phys. J. C } {\bf 2018}, {\em 78}, 825.

\bibitem[Xu(2018)]{Xu2018}
Xu, H.; Ma, B.-Q.
Regularity of high energy photon events from gamma ray bursts.
{\em J. Cosmol. Astropart. Phys.} {\bf 2018}, {\em  {01}
}, 050.

\bibitem[Huang(2018)]{Huang2018}
Huang, Y.; Ma, B.-Q.
Lorentz violation from gamma-ray burst neutrinos.
{\em Commun. Phys.} {\bf 2018}, {\em 1}, 62. 
\url{https://doi.org/10.1038/s42005-018-0061-0}.

\bibitem[Huang(2019)]{Huang2019}
Huang, Y.; Li, H.; Ma, B.-Q.
Consistent Lorentz violation features from near-TeV IceCube neutrinos.
{\em Phys. Rev. D} {\bf 2019}, {\em 99}, 123018. 


\bibitem[Li(2020)]{Li2020}
Li, H.; Ma, B.-Q.
Light speed variation from active galactic nuclei.
{\em Sci. Bull.} {\bf 2020}, {\em 65}, 262–266.

\bibitem[Pan(2020)]{Pan2020}
Pan, Y.; Qi, J.; Cao, S.; Liu, T.; Liu, Y.; Geng, S.; Lian, Y.; Zhu, Z.-H.
Model-independent constraints on Lorentz invariance violation: Implication from updated gamma-ray burst observations.
{\em Astrophys. J.} {\bf 2020}, {\em 890}, 169.

\bibitem[Acciari(2020)]{Acciari2020}
Acciari, V.A.; Ansoldi, S.; Antonelli, L.A.; Arbet Engels, A.; Baack, D.; Babić, A.; Banerjee, B.; Barres de Almeida, U.; Barrio, J.A.; Becerra González, J.; et~al.
 Bounds on Lorentz invariance violation from MAGIC observation of GRB 190114C.
{\em Phys. Rev. Lett.} {\bf 2020}, {\em 125}, 021301.

\bibitem[Chen(2021)]{Chen2021}
Chen, Y.; Ma, B.-Q.
Novel pre-burst stage of gamma-ray bursts from machine learning. 
{\em J. High Energy Astrophys.} {\bf 2021}, {\em 32}, 78--86.

\bibitem[JieZ(2021)]{JieZ2021-1}
Zhu, J.; Ma, B.-Q.
Pre-burst events of gamma-ray bursts with light speed variation.
{\em Phys. Lett. B} {\bf 2021}, {\em 820}, 136518.

\bibitem[JieZ(2021)]{JieZ2021-2}
Zhu, J.; Ma, B.-Q.
Pre-burst neutrinos of gamma-ray bursters accompanied by high-energy photons.
{\em Phys. Lett. B} {\bf 2021}, {\em 820}, 136546.








\bibitem[Jacob(2008)]{Jacob2008}
Jacob, U.; Piran, T.
Lorentz-violation-induced arrival delays of cosmological particles.
{\em J. Cosmol. Astropart. Phys.} {\bf 2008}, {\em  {01}}, 031.

\bibitem[Pfeifer(2018)]{Pfeifer2018}
Pfeifer, C.
Redshift and lateshift from homogeneous and isotropic modified dispersion relations.
{\em Phys. Lett. B} {\bf 2018}, {\em 780}, 246--250.


\bibitem[JieZ(2023)]{JieZ-unp}
Zhu, J.; Ma, B.-Q.
Trajectories of astroparticles in pseudo-Finsler spacetime with the most general modified dispersion. 
\textit{Eur. Phys. J. C} {\bf 2023}, {\em 83}, 349.
\url{https://doi.org/10.1140/epjc/s10052-023-11517-8}


\bibitem[Judes(2003)]{Judes2003}
Judes, S.
Conservation laws in ‘‘doubly special relativity’’. 
{\em Phys. Rev. D} {\bf 2003}, {\em 68}, 045001.

\bibitem[Lukierski(2003)]{Lukierski2003}
Lukierski, J.; Nowicki, A.
Doubly spacial relativity versus $\kappa$-deformation of relativistic kinematics.
{\em Int. J. Mod. Phys. A} {\bf 2003}, {\em 18}, 7--18.


\bibitem[Gubitosi(2013)]{Gubitosi2013}
Gubitosi, G.; Mercati, F.
Relative Locality in $\kappa$-Poincaré.
{\em Class. Quant. Grav.} {\bf 2013}, {\em 30}, 145002.













\bibitem[Lukierski(1991)]{Lukierski1991-1}
Lukierski, J; Ruegg, H.; Nowickl, A.; Tolstoy, V.N.
Q deformation of Poincare algebra. 
{\em Phys. Lett. B} {\bf 1991}, {\em 264},  {331--338.} 


\bibitem[Lukierski(1991)]{Lukierski1991-2}
Lukierski, J; Nowicki, A.; Ruegg, H.
Real forms of complex quantum anti-De Sitter algebra $U_q(Sp(4:C))$ and their contraction schemes. 
{\em Phys. Lett. B} {\bf 1991}, {\em 271}, 321.

\bibitem[Majid(1994)]{Majid1994}
Majid, S.; Ruegg, H.
Bicrossproduct structure of kappa Poincare group and noncommutative geometry. 
{\em Phys. Lett. B} {\bf 1994}, {\em 334}, 348.

\bibitem[Lukierski(1995)]{Lukierski1995}
\textls[-15]{Lukierski, J; Ruegg, H.; Zakrzewski, W.J.
Classical quantum mechanics of free kappa relativistic systems. 
{\em Annals Phys.} {\bf 1995}, {\em 243}, 90.}

\bibitem[Kowalski(2001)]{Kowalski2001}
Kowalski-Glikman, J.
Observer independent quantum of mass. 
{\em Phys. Lett. A} {\bf 2001}, {\em 286}, 391.

\bibitem[Bruno(2001)]{Bruno2001}
Bruno, N.R.; Amelino-Camelia, G.; Kowalski-Glikman, J.
Deformed boost transformations that saturate at the Planck scale. 
{\em Phys. Lett. B} {\bf 2001}, {\em 522}, 133.

\bibitem[Kowalski(2002)]{Kowalski2002}
 Kowalski-Glikman, J.; Nowak, S.
Doubly special relativity theories as different bases of $\kappa$-Poincaré algebra.
{\em Phys. Lett. B} {\bf 2002}, {\em 539}, 126--132.


\bibitem[Mignemi(2007)]{Mignemi2007}
Mignemi, S.
Doubly special relativity and Finsler geometry. 
{\em Phys. Rev. D} {\bf 2007}, {\em 76}, 047702.













\bibitem[ Kostelecký(2011)]{finsler-sme}
Kostelecký, V.A. 
Riemann–Finsler geometry and Lorentz-violating kinematics. 
{\em Phys. Lett. B} {\bf 2011}, {\em 701}, 137--143.

\bibitem[ Kostelecký(2012)]{bipartite-sme}
Kostelecký, V.A.; Russell, N.; Tso, R.
Bipartite Riemann–Finsler geometry and Lorentz violation. 
{\em Phys. Lett. B} {\bf 2012}, {\em 716}, 470--474.






\bibitem[Colladay(2015)]{Colladay2015}
Colladay, D.; McDonald, P.
Singular Lorentz-violating Lagrangians and associated Finsler structures.
{\em Phys. Rev. D} {\bf 2015}, {\em 92}, 085031.

\bibitem[Russell(2015)]{Russell2015}
Russell, N.
Finsler-like structures from Lorentz-breaking classical particles.
{\em Phys. Rev. D} {\bf 2015}, {\em 91}, 045008.

\bibitem[Schreck(2016)]{Schreck2016}
Colladay, M.
Classical Lagrangians and Finsler structures for the nonminimal fermion sector of the standard model extension.
{\em Phys. Rev. D} {\bf 2016}, {\em 93}, 105017.

\bibitem[Edwards(2018)]{finsler-scalar}
Edwards, B.R.; Kostelecký, V.A.
Riemann–Finsler geometry and Lorentz-violating scalar fields.
{\em Phys. Lett. B} {\bf 2018}, {\em 786}, 319--326.

\bibitem[ Schreck(2019)]{sme-higher-order1}
 Schreck, M.
 Classical Lagrangians for the nonminimal Standard-Model Extension at higher orders in Lorentz violation.
 {\em Phys. Lett. B} {\bf 2019}, {\em 793}, 70--77.

 \bibitem[Schreck(2021)]{sme-higher-order2}
 Reis, J.A.A.S.; Schreck, M.
 Classical Lagrangians for the nonminimal spin-nondegenerate Standard-Model extension at higher orders in Lorentz violation.
 {\em Phys. Rev. D} {\bf 2021}, {\em 103}, 095029.

\bibitem[ Kostelecký(2001)]{sclcpt}
Kostelecký, V.A.; Lehnert, R. 
Stability, causality, and Lorentz and $CPT$ violation. 
{\em Phys. Rev. D} {\bf 2001}, {\em 63}, 065008.

\bibitem[Kostelecký(2010)]{kinematics-lv}
Kostelecký, V.A.; Russell, N. 
Classical kinematics for Lorentz violation. 
{\em Phys. Lett. B} {\bf 2010}, {\em 693}, 443.

 \bibitem[ Silva(2014)]{Silva2014}
Silva, J.E.G.; Almeida, C.A.S.
Kinematics and dynamics in a bipartite-Finsler spacetime. 
{\em Phys. Lett. B} {\bf 2014}, {\em 731}, 74--79.

\bibitem[ Colladay(2017)]{Colladay2017}
Colladay, D.
Extended hamiltonian formalism and Lorentz-violating lagrangians. 
{\em Phys. Lett. B} {\bf 2017}, {\em 772}, 694--698.

\bibitem[ Silva(2019)]{Silva2019}
Silva, J.E.G.; Maluf, R.V.; Almeida, C.A.S.
Bipartite-Finsler symmetries. 
{\em Phys. Lett. B} {\bf 2019}, {\em 798}, 135009.




\bibitem[Bogoslovsky(1977)]{Bogoslovsky1977}
Bogoslovsky, G.Y.
A special-relativistic theory of the locally anisotropic space-time. 
{\em Nuov. Cim. B} {\bf 1977}, {\em 40}, 99--115.

\bibitem[Bogoslovsky(1998)]{Bogoslovsky1998}
Bogoslovsky, G.Y.; Goenner, H.F.
On a possibility of phase transitions in the geometric structure of space-time. 
{\em Phys. Lett. A} {\bf 1998}, {\em 244}, 222--228.

\bibitem[Bogoslovsky(1999)]{Bogoslovsky1999-1}
Bogoslovsky, G.Y.; Goenner, H.F.
Finslerian spaces possessing local relativistic symmetry. 
{\em Gen. Relativ. Gravit.} {\bf 1999}, {\em 31}, 1565.

\bibitem[Bogoslovsky(1999)]{Bogoslovsky1999-2}
Goenner, H.F.; Bogoslovsky, G.Y.
A class of anisotropic (Finsler) space-time geometries. 
{\em Gen. Relativ. Gravit.} {\bf 1999}, {\em 31}, 1383.


\bibitem[Ahluwalia(2010)]{Ahluwalia2010}
Ahluwalia, D.; Horvath, S.
Very special relativity as relativity of dark matter: The Elko connection.
{\em  {J. High Energy Phys.}} {\bf 2010}, {\em  {1011}
}, 078.

\bibitem[Kogut(1970)]{Kogut1970}
Kogut, J.B.; Soper, D.E.
Quantum Electrodynamics in the Infinite-Momentum Frame.
{\em Phys. Rev. D} {\bf 1970}, {\em 1}, 2901.

\bibitem[Gibbons(2007)]{vsr-finsler}
Gibbons, G.W.; Gomis, J.; Pope, C.N.
General very special relativity is Finsler geometry.
{\em Phys. Rev. D} {\bf 2007}, {\em 76}, 081701.

\bibitem[Monique(1967)]{Monique1967}
Levy‐Nahas, M.
Deformation and Contraction of Lie Algebras.
{\em J. Math. Phys.} {\bf 1967}, {\em 8}, 1211.

\bibitem[Glashow(2006)]{vsr-neutrino}
Cohen, A.G.; Glashow, S.L.
A Lorentz-Violating Origin of Neutrino Mass? \emph{ {arXiv}} \textbf{ {2006}},
 	arXiv:hep-ph/0605036. 


 \bibitem[Cheon(2009)]{Cheon2009}
 Cheon, S.; Lee, C.; Lee, S.J.
 $SIM(2)$-invariant modifications of electrodynamic theory.
{\em Phys. Lett. B} {\bf 2009}, {\em 679}, 73.

\bibitem[Alfaro(2013)]{Alfaro2013}
Alfaro, J.
Non-Abelian fields in very special relativity.
{\em Phys. Rev. D} {\bf 2013}, {\em 88}, 085023.

\bibitem[Alfaro(2014)]{Alfaro2014}
Alfaro, J.; Rivelles, V.O.
Very special relativity and Lorentz violating theories.
{\em Phys. Lett. B} {\bf 2014}, {\em 734}, 239--244.

\bibitem[Alfaro(2015)]{Alfaro2015}
Alfaro, J.; González, P.; Avila, R.
Electroweak standard model with very special relativity.
{\em Phys. Rev. D} {\bf 2015}, {\em 91}, 105007.

\bibitem[Alfaro(2017)]{Alfaro2017}
Alfaro, J.
A $Sim(2)$ invariant dimensional regularization.
{\em Phys. Lett. B} {\bf 2017}, {\em 772}, 100--104.

\bibitem[Alfaro(2018)]{Alfaro2018}
Alfaro, J.
Loop Corrections in Very Special Relativity Standard Model.
{\em J. Phys. Conf. Ser.} {\bf 2018}, {\em 952}, 012009.

\bibitem[Alfaro(2019)]{Alfaro2019}
Alfaro, J.
Feynman Rules, Ward Identities and Loop Corrections in Very Special Relativity Standard Model.
{\em Universe} {\bf 2019}, {\em 5}, 16.

\bibitem[Alfaro(2019)]{Alfaro2019-2}
Alfaro, J.; Soto, A.
Photon mass in very special relativity.
{\em Phys. Rev. D} {\bf 2019}, {\em 100}, 055029.

\bibitem[Alfaro(2019)]{Alfaro2019-3}
Alfaro, J.; Soto, A.
Schwinger model à la Very Special Relativity.
{\em Phys. Lett. B} {\bf 2019}, {\em 797}, 134923.

\bibitem[Alfaro(2021)]{Alfaro2021}
Alfaro, J.
Axial anomaly in very special relativity.
{\em Phys. Rev. D} {\bf 2021}, {\em 103}, 075011.

\bibitem[Alfaro(2022)]{Alfaro2022}
Alfaro, J.; Santoni, A.
Very special linear gravity: A gauge-invariant graviton mass.
{\em Phys. Lett. B} {\bf 2022}, {\em 829}, 137080.

\bibitem[Haghgouyan(2022)]{Haghgouyan2022}
Haghgouyan, Z.; Ghasemkhani, M.; Bufalo, R.; Soto, A.
Induced non-Abelian Chern–Simons effective action in very special relativity.
{\em Eur. Phys. J. Plus} {\bf 2022}, {\em 137}, 732.

\bibitem[Bufalo(2022)]{Bufalo2022}
Bufalo, R.; Ghasemkhani, M.
Path integral analysis of the axial anomaly in Very Special Relativity.
{\em Mod. Phys. Lett. A} {\bf 2022}, {\em 37}, 2250002.

\bibitem[Cheng-Yang(2015)]{Cheng-Yang2015}
Lee, C.
Quantum field theory with a preferred direction: The very special relativity framework.
{\em Phys. Lett. B} {\bf 2016}, {\em 93}, 045011.

\bibitem[Ilderton(2016)]{Ilderton2016}
Ilderton, A.
Very Special Relativity as a background field theory.
{\em Phys. Rev. D} {\bf 2016}, {\em 94}, 045019.

\bibitem[Bogoslovsky(2004)]{Bogoslovsky2004}
Bogoslovsky, G.Y.; Goenner, H.F.
Concerning the generalized Lorentz symmetry and the generalization of the Dirac equation.
{\em Phys. Lett. A} {\bf 2004}, {\em 323}, 40.

\bibitem[Alvarez(2008)]{Alvarez2008}
Alvarez, E.; Vidal, R.
Very special (de Sitter) relativity.
{\em Phys. Rev. D} {\bf 2008}, {\em 77}, 127702.

\bibitem[Mück(2008)]{Muck2008}
Mück, W.
Very special relativity in curved space–times.
{\em Phys. Lett. B} {\bf 2008}, {\em 670}, 95--98.

\bibitem[Kouretsis(2009)]{Kouretsis2009}
Kouretsis, A.P.; Stathakopoulos, M.; Stavrinos, P.C.
General very special relativity in Finsler cosmology.
{\em Phys. Rev. D} {\bf 2009}, {\em 79}, 104011.

\bibitem[Fuster(2018)]{Fuster2018}
Fuster, A.; Pabst, C.; Pfeifer, C.
Berwald spacetimes and very special relativity.
{\em Phys. Rev. D} {\bf 2018}, {\em 98}, 084062.








\end{thebibliography}
\end{document}